\documentclass[12pt,preprint]{aastex}
\newcommand{\lam}{$\lambda$}

\newcommand{\ecss}{erg cm$^{-2}$ s$^{-1}$ sr$^{-1}$} % erg/cm2/s/sr
\newcommand{\kms}{km~s$^{-1}$}

\newcommand{\vtl}{$\vartriangleleft$}
\newcommand{\btl}{$\blacktriangleleft$}
\renewcommand{\ion}[2]{#1\,{\sc #2}}
\newcommand{\tblll}{4}
\newcommand{\figdem}{8}
\newcommand{\figim}{1}
\newcommand{\tclass}{1}

\def\ionx[#1 #2]{#1\,{\sc #2}}

\begin{document}


\title{CHIANTI -- An atomic database for emission lines. XI. EUV
  emission lines of Fe\,VII, Fe\,VIII and Fe\,IX
  observed by Hinode/EIS.}


\author{P. R. Young\altaffilmark{1,2}
\and E. Landi\altaffilmark{2}}

\altaffiltext{1}{George Mason University, 4400 University Drive, Fairfax, VA 22030}
\altaffiltext{2}{Naval Research Laboratory, Space Science Division, Washington, DC 20375}


\begin{abstract}
A detailed study of emission lines from \ion{Fe}{vii},
\ion{Fe}{viii} and \ion{Fe}{ix} observed by the EUV Imaging
Spectrometer on board the Hinode satellite is presented. Spectra in
the ranges 170--212~\AA\ and 246--292~\AA\ show strongly enhanced
lines from the upper solar transition region (temperatures $5.4\le
\log\,T \le 5.9$) allowing a number of new line identifications to be
made. Comparisons of \ion{Fe}{vii} lines with predictions from a new
atomic model reveal new plasma diagnostics, however there
are a number of disagreements between theory and observation for
emission line ratios insensitive to density and temperature,
suggesting improved atomic data are required. Line ratios for
\ion{Fe}{viii} also show discrepancies with theory, with the strong
\lam185.21 and \lam186.60 lines under-estimated by 60--80~\%\ compared
to lines between 192 and 198~\AA. A newly-identified multiplet between
253.9 and 255.8~\AA\ offers excellent temperature diagnostic
opportunities relative to the lines between 185--198~\AA, however the
atomic model under-estimates the strength of these lines by factors
3--6. Two new line identifications are made for \ion{Fe}{ix} at
wavelengths 176.959~\AA\ and 177.594~\AA, while seven other lines
between 186 and 200~\AA\ are suggested to be due to \ion{Fe}{ix} but
for which transition identifications can not be made.
The new atomic data for
\ion{Fe}{vii} and \ion{Fe}{ix} are demonstrated to significantly
modify models for the response function of the TRACE 195~\AA\ imaging
channel, affecting temperature determinations from this channel. The
data will also affect the response functions for other solar EUV
imaging instruments such as SOHO/EIT, STEREO/EUVI and the upcoming
AIA instrument on the Solar Dynamics Observatory.
\end{abstract}

\keywords{line: identification --- atomic data --- Sun: corona --- Sun:
  UV radiation --- Sun: transition region} 

\section{Introduction}

The iron ions are extremely important for the study of the solar
corona as they give rise to many strong emission lines in the extreme
ultraviolet wavelength range 90--400~\AA\ that have been exploited
for over 40 years \citep{gabriel66,phillips08}. Resonance lines of species such
\ion{Fe}{ix}, \ion{Fe}{xii} and \ion{Fe}{xv} have been selected for
observation by EUV imaging instruments such as the EUV Imaging
Telescope (EIT) on SOHO \citep{delab95}, the Transition Region and Coronal Explorer
\citep[TRACE,][]{handy99}, and  the Extreme UltraViolet Imagers
\citep[EUVI,][]{howard08} on the twin  
STEREO spacecraft. Spectroscopically, the iron lines have been
measured by a range of space instrumentation, from the early Orbiting
Solar Observatories (OSOs), through to the Skylab S082A spectrometer,
the Coronal Diagnostic Spectrometer (CDS) on board SOHO
\citep{harrison95} and, most
recently, the Hinode/EIS instrument \citep{culhane07}.
The complexity of the iron ions' atomic structure
leads to many emission line pairs that are sensitive to electron
density and, indeed, iron ions form most of the best coronal density
diagnostics. 

The iron ions \ion{Fe}{x} to \ion{Fe}{xiv} have received a lot of
attention from observers and atomic physicists as they give rise to
many strong lines in the solar spectrum. \ion{Fe}{vii--ix}, by comparison,
have few strong lines and atomic calculations are less
common. \ion{Fe}{ix} has a single 
very strong emission line at 171.07~\AA\ and a density diagnostic
involving two lines at 241.74 and 244.91~\AA, but few other lines had
been identified in the solar spectrum until \citet{young09} presented four
new line identifications based on Hinode/EIS spectra. \ion{Fe}{viii}
gives rise to much weaker lines than the \ion{Fe}{ix--xiv} ions and
none of them have diagnostic potential, so the  ion has received
little attention. However, 
\ion{Fe}{viii} can make a significant contribution to the 195~\AA\
imaging channel of the SOHO/EIT and TRACE instruments for polar plumes
\citep{delzanna03} 
and coronal loops \citep{delzanna03b}, while the high sensitivity of the Hinode/EIS
spectrometer has led to \ion{Fe}{viii} lines being observed regularly
\citep{young07a}. 
\ion{Fe}{vii} is predicted to form around $\log\,T=5.4$
\citep{bryans09} where the emission measure of typical coronal
features is much lower than for the higher temperature iron ions
\citep[e.g.,][]{raymond81}. In addition, the atomic structure for
\ion{Fe}{vii} produces a large number of emission lines of relatively
weak strength rather than a few strong lines such as found for the
higher ionization stages of iron. These facts combine to make the
observed \ion{Fe}{vii} solar spectrum very weak compared to the other
iron ions and so it has been little studied. 

The present work considers the lines of \ion{Fe}{vii},
\ion{Fe}{viii} and \ion{Fe}{ix} measured in the Hinode/EIS spectrum presented by
\citet[][hereafter Paper~I]{landi09},
%Landi et al.~(2009, hereafter Paper~I),
comparing measured intensities with predictions from
atomic data in the CHIANTI atomic database \citep{dere97,dere09} and
investigating their diagnostic potential. The
observed spectrum shows \ion{Fe}{vii--ix} lines that are strongly enhanced
over typical quiet Sun and active region conditions, and thus is ideal
for studying lines that are normally too weak to be observed. In
addition we discuss the formation temperatures of
\ion{Fe}{vii} and \ion{Fe}{viii} which appear to be discrepant with other ions. 

\section{Observations and data reduction}

The Hinode/EIS data-set and data reduction method were described in
detail in Paper~I and are only summarized briefly here. The
observation took place on 2007 February 21 at 01:15~UT, when the
footpoint regions of active region AR~10942 were observed. Complete
EIS spectra were obtained, and calibration was performed with the
standard EIS routine EIS\_PREP. Spectra from a strong brightening near
the base of some coronal loops were averaged, taking care to adjust
for spatial 
offsets between images obtained at different wavelengths. A complete
line list from the final, averaged spectrum was presented in Paper I
and a differential emission measure constructed. 
%and
%analysed the 
Emission line strengths from all ions formed in the range
$5.4\le \log\,T\le 6.0$ were analysed, except for the iron ions
\ion{Fe}{vii--ix} which  are studied in the present work.

In the following sections we will refer to the differential emission
measure (DEM) curve derived in Paper~I, which was displayed in
Fig.~\figdem\ of that work. In addition we also refer to the intensity
images formed in various lines that are often valuable for
determining the emitting species of unidentified lines or classifying
blends. Fig.~\figim\ of Paper~I shows images from a wide range of ions
belonging to different temperatures. The term ``class'' is used to
group together emission lines for which the image morphology is
similar to one of these reference images. For example class~C lines
are similar to \ion{Fe}{vii} \lam195.39, class~D lines similar to
\ion{Fe}{viii} \lam185.21, etc. The complete list of temperature
classes is given in Table~\tclass\ in Paper~I.

\section{Ion fraction comparison}\label{sect.ion-fraction}

The abundance of the emitting ion is a key ingredient in the analysis
of the emission of spectral lines. Ion abundance calculations rely on
ionization and  recombination rates that are continuously updated and
improved both by laboratory  measurements and by new {\em ab initio}
calculations. Several data sets have been  made available in the
literature to provide reliable ion abundances as a function  of
temperature, a few of them published very recently. 

\begin{figure}
\includegraphics[width=12.0cm,height=17.0cm,angle=90]{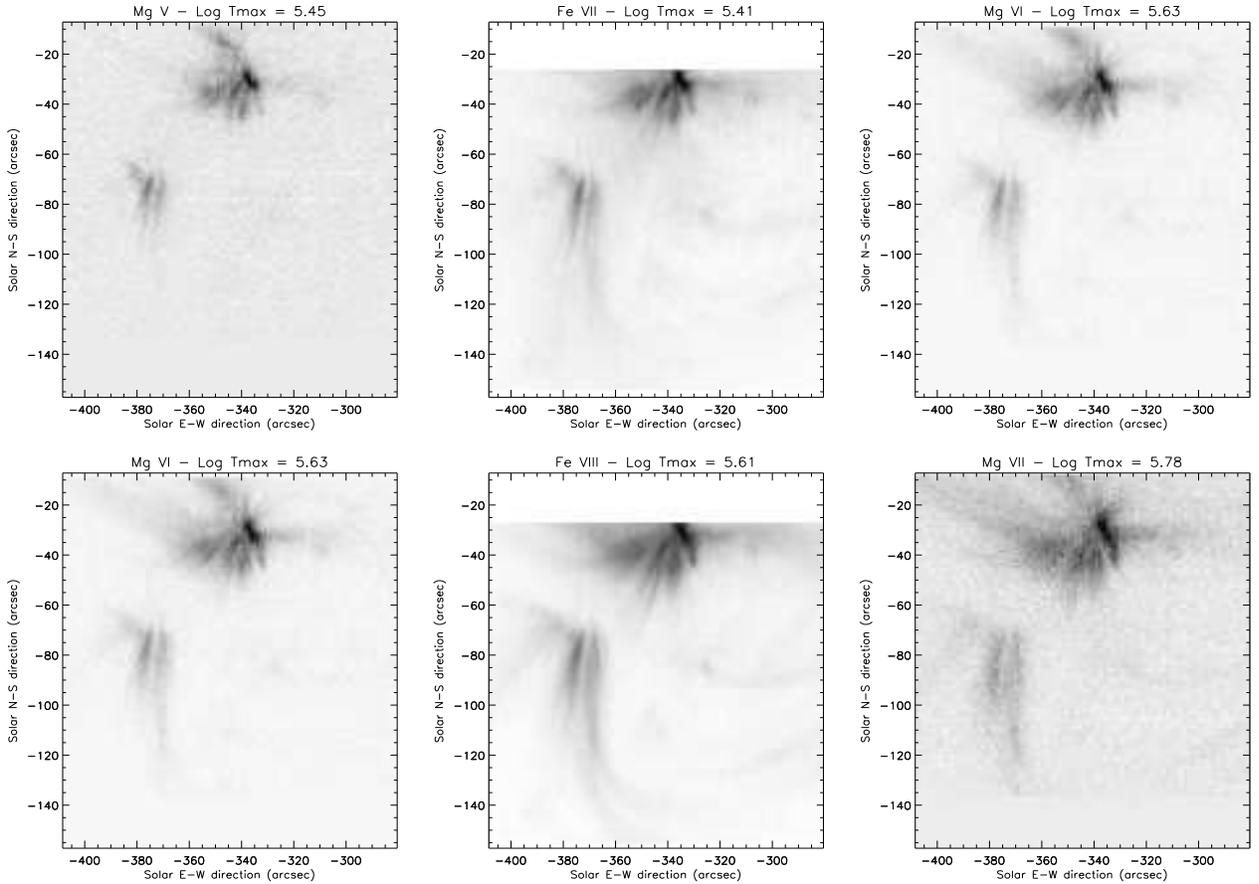}
\caption{\label{ion_frac_1} Intensity maps of a few selected ions. $\log T_{\rm max}$
is the temperature of maximum abundance of each ion, according to \cite{mazzotta98}.}
\end{figure}

\cite{young07b} noted that the temperature of maximum abundance,
$T_{\rm max}$, of  \ionx[Fe viii] predicted by \citet{mazzotta98}
appeared to be too
low compared to observations. By comparing images obtained in several
emission lines the authors demonstrated that the intensity
distribution of \ion{Fe}{viii}, predicted to be found at $\log T_{\rm max}
= 5.56$, was very similar to the intensity image of \ionx[Si vii]
($\log T_{\rm max} = 5.76$). Since, in terms of atomic structure,
\ion{Si}{vii} is a much simpler ion than \ion{Fe}{viii},
\citet{young07b} suggested the \citet{mazzotta98} ion fractions for
\ion{Fe}{viii} were incorrect. Using the observations from the 2007
February 21 dataset we demonstrate that a similar effect is found for
\ion{Fe}{vii}. Figure~\ref{ion_frac_1} shows the intensity maps of the emitting region obtained 
with lines from \ionx[Mg v-vii] and \ionx[Fe vii-viii]. The top row compares
a \ionx[Fe vii] image with images from \ionx[Mg v-vi], and the bottom
row compares 
\ionx[Fe viii] with \ionx[Mg vi-vii]. The intensity maps show that \ionx[Fe vii]
and \ionx[Mg vi] images are very similar, implying similar $T_{\rm
  max}$ values, while 
\ionx[Fe viii] and 
\ionx[Mg vii] are also very similar. The \citet{mazzotta98}
calculations, however,
place  \ionx[Fe vii] closer to \ionx[Mg v] and 
\ionx[Fe viii] closer to \ionx[Mg vi] (see Table~\ref{ion_frac_0}).

There have been several ion balance calculations performed over the
past quarter of a century, and the \citet{mazzotta98} work is the most
commonly used of recent years. Table~\ref{ion_frac_0} compares $T_{\rm
  max}$ values for \ionx[Mg vi-vii] and \ionx[Fe vii-viii] from these
calculations, including three recent works
\citep{bryans06,bryans09,dere09}. The values for \ionx[Mg vi-vii] are
remarkably consistent, however a marked decrease is found for both
\ion{Fe}{vii} and \ion{Fe}{viii} from the \citet{raymond92}
calculations onwards. Interestingly, the older $T_{\rm max}$ values of
\citet{shull82} and \citet{rothenflug85} for
\ion{Fe}{vii} and \ion{Fe}{viii} are more consistent with the
intensity distributions shown in Figure~\ref{ion_frac_1}.

\begin{figure}
\begin{center}
\includegraphics[width=7.0cm,height=12.0cm,angle=90]{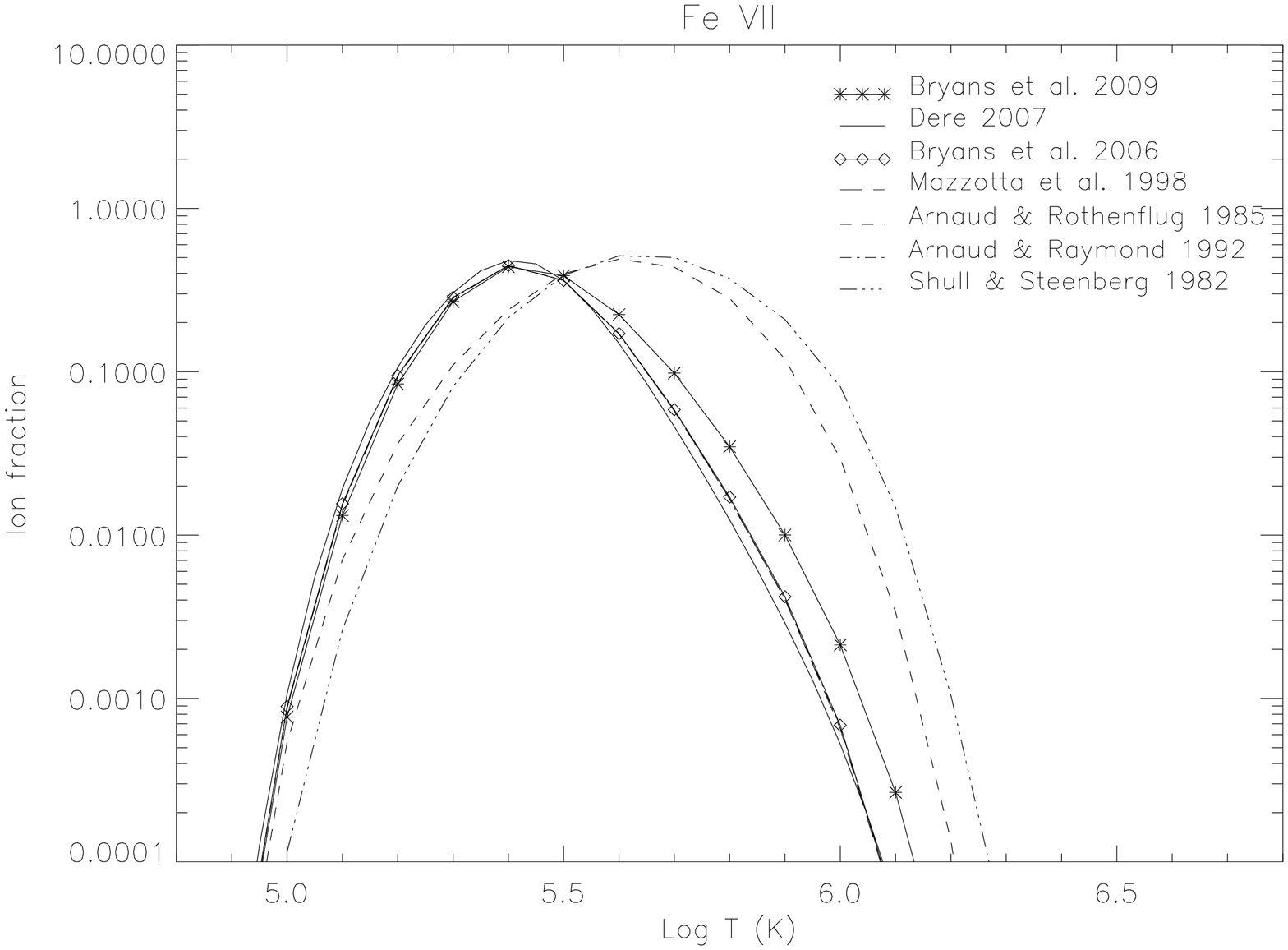}
\includegraphics[width=7.0cm,height=12.0cm,angle=90]{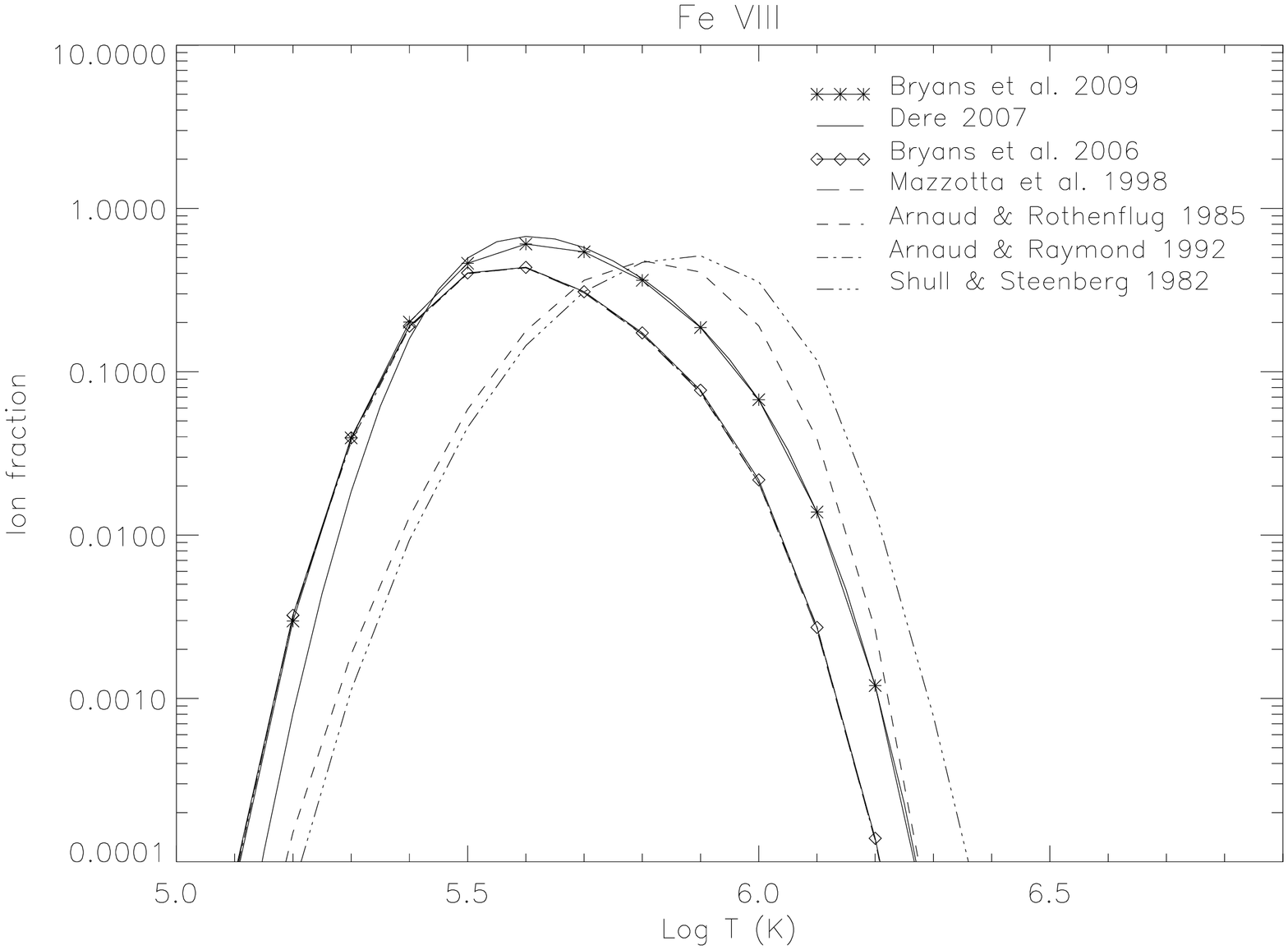}
\includegraphics[width=7.0cm,height=12.0cm,angle=90]{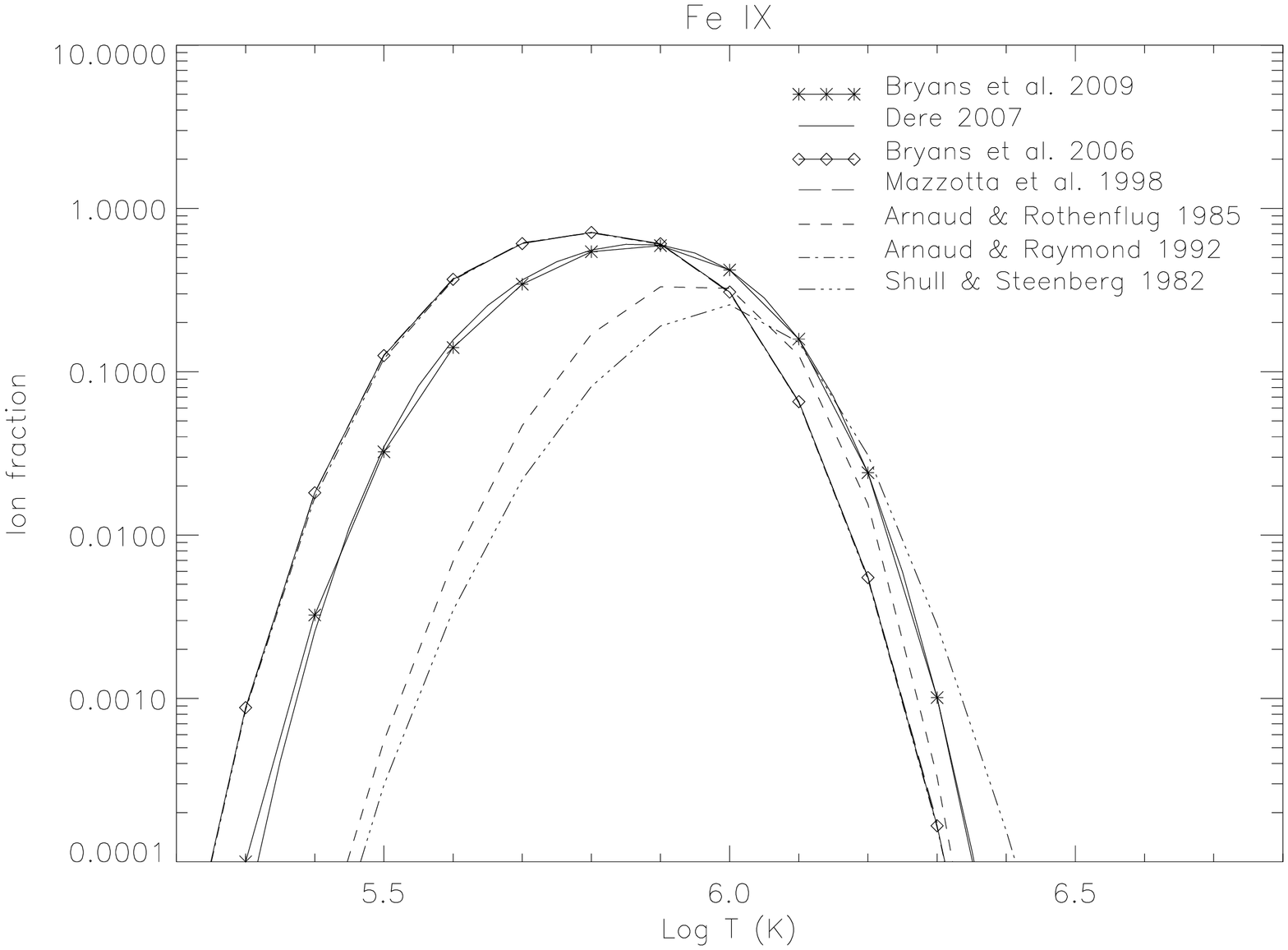}
\end{center}
\caption{\label{ion_frac_2} Comparison of ion fraction datasets for \ionx[Fe vii], \ionx[Fe viii]
and \ionx[Fe ix], using the most recent calculations.}
\end{figure}

In Fig.~\ref{ion_frac_2} we compare the ion fractions for \ionx[Fe
vii-ix] 
as a function of temperature. Significant differences are found
between the curves, but the largest ones take place with the iron ion update of 
\cite{raymond92}, that resulted in differences of up to one order of magnitude 
from the previous calculations.
Investigating the reasons for this lies beyond the
scope of the present paper, but there are two main avenues to be considered:
1) the rates used for \ionx[Fe vii-ix] are incorrect, or 2) some other process
(e.g. dynamics, density effects) may be influencing the \ionx[Fe vii-ix] ion fractions.

\begin{deluxetable}{lcccccc}
\tablecaption{Values of the maximum abundance temperature $\log T_{\rm max}$
(in K) for \ionx[Fe vii-viii] and \ionx[Mg v-vii], from the seven most recent calculations
available in the literature.\label{ion_frac_0}}
\tablehead{Data set & \ionx[Mg v] & \ionx[Fe vii] & \ionx[Mg vi] & \ionx[Fe viii] & \ionx[Mg vii] }
\tablewidth{0pt}
\startdata
\cite{shull82}      & 5.44 & 5.64 & 5.63 & 5.88 & 5.81 \\
\cite{rothenflug85} & 5.43 & 5.61 & 5.63 & 5.82 & 5.80 \\
\cite{raymond92}    & 5.43 & 5.42 & 5.63 & 5.57 & 5.80 \\
\cite{mazzotta98}   & 5.43 & 5.42 & 5.64 & 5.56 & 5.80 \\
\cite{bryans06}     & 5.44 & 5.42 & 5.62 & 5.56 & 5.78 \\
\cite{dere09}       & 5.45 & 5.41 & 5.63 & 5.61 & 5.78 \\
\cite{bryans09}     & 5.45 & 5.42 & 5.63 & 5.62 & 5.78 \\
\enddata
\end{deluxetable}

\section{Fe\,VII}\label{sect.fe7}

\ion{Fe}{vii} lines in the wavelength range 170--300~\AA\ have
received little  attention from solar spectroscopists as the lines are
generally very weak and there have been no atomic data available to
model the line intensities.  However, two extensive atomic calculations have
been published in recent years that allow the lines to be modeled,
while the launch of EIS in 2006 has allowed the routine measurement of
parts of the 170-300~\AA\ window at high resolution and high
sensitivity. The spectrum presented in Paper~I shows strongly enhanced lines of
\ion{Fe}{vii} compared to normal quiet Sun or active region spectra
and so presents an excellent opportunity for comparing theory with
observations.

Atomic data suitable for modeling electron-excited emission lines have
been published by \citet{zeng05} and \citet{witthoeft08} and these
are the first calculations that yield data for the excited
configurations of the ion that give rise to the emission lines in the
extreme ultraviolet. The \citet{zeng05} calculations were
performed using the Flexible Atomic Code \citep[FAC;][]{gu03} which
uses the distorted wave approximation, while \citet{witthoeft08}
performed a R-matrix calculation using the intermediate-coupling frame
transformation (ICFT) method of \citet{griffin98}. Generally R-matrix
calculations are considered superior to distorted wave as they
allow the resonant enhancement of collision strengths to be accurately
modeled, although for excited configurations and high energies the
results should be similar. \citet{witthoeft08} made a comparison with
\citet{zeng05} and found good agreement for both radiative decay rates
and collision strengths. For this work we choose to use the
\citet{witthoeft08} for our atomic model. 

The \citet{witthoeft08} data have been put into the format used by the
CHIANTI database, 
and will be made publicly available in a future version of the
database. For inclusion in CHIANTI the collision strengths have been
fit with splines using the method outlined in \citet{dere97}, while
experimental energy levels have been taken from \citet{ekberg81},
\citet{ekberg03} and version 3 of the NIST 
database \citep{ralchenko08}. Many levels do not have experimental energies and for these
the theoretical values of  \citet{witthoeft08} were used. Some new and
revised energy values are suggested from the present work and will be
discussed below.

Since the atomic model constructed here is the first to yield
intensity predictions for lines from excited configurations of
\ion{Fe}{vii} at typical coronal densities and temperatures, we
outline briefly some of the properties of the lines in terms of
diagnostic potential. Firstly, above densities of $10^9$~cm$^{-3}$ the
nine ground configuration levels of \ion{Fe}{vii} are in a
quasi-Boltzmann distribution, and thus relative populations change
little relative to each other above this value. This means that emission line ratios show little
density sensitivity at typical coronal densities of
$10^9$--$10^{13}$~cm$^{-3}$. There is density sensitivity amongst some
emission lines below
$10^9$~cm$^{-3}$ which may be useful when studying 
coronal hole regions and specific diagnostics are highlighted in the
sections below.
The lines formed in the 170--300~\AA\ range have a wide range of
excitation potentials and so they can show significant
temperature sensitivity.
In the sections below we will highlight some useful temperature
diagnostics and compare results from these ratios. 

Ratios
that are insensitive to the physical conditions of the atmosphere are
also valuable as a check on the atomic physics parameters. In order to
compare observed ratios with theory in these cases, we compute the
theoretical ratio over the temperature range $\log\,T=5.4$--5.7 at
0.05~dex intervals, and the density range $\log\,N_{\rm e}=8.0$--10.0
at 0.2~dex intervals. The listed theoretical ratio is then given 
as the average ratio over these ranges, and the ``error'' is the
2-$\sigma$ value of the ratio over the ranges
(Table~\ref{tbl.fe7.insens}).  The temperature range 
has been chosen as $\log\,T=5.4$--5.7 based on the discussion in
Sect.~\ref{sect.ion-fraction}.
The density
range chosen is based on the density measured from the \ion{Mg}{vii}
\lam280.75/\lam278.39 diagnostic (see Paper~I) and allowing for up to an
order of magnitude variation from this value. 

Table~\tblll\ in Paper~I listed line intensities predicted from
the differential emission measure (DEM) curve and for most of the
\ion{Fe}{vii} lines the predictions are lower than the observed
intensities by factors of 2--5. The \ion{Fe}{vii} lines were not
included when deriving the DEM curve since no previous check of the
\ion{Fe}{vii} atomic data has been performed. We believe the large
underestimates of the intensities principally arise from the
inaccurate ion fraction curve for \ion{Fe}{vii} discussed in the
previous section. The DEM predicted intensities include a convolution of
the theoretical ion fraction curve and the DEM, and Figure~8 of
Paper~I shows that the DEM has a rather low value at $\log\,T=5.4$,
the \ion{Fe}{vii} $T_{\rm max}$ value of \citet{dere09}. If the ion
fraction curve was shifted to around $\log\,T=5.6$--5.7, then it would
sample larger values of the DEM and so increased predicted intensities
would result. In the \ion{Fe}{vii} discussions that follow we will
generally not refer to the DEM intensity predictions for this reason.

Before embarking on our analysis of the observed EIS lines we first
consider the measured EIS wavelengths of those transitions that have
been identified previously. The line list of \citet{ekberg81} is the
only comprehensive one in the literature for the 170--300~\AA\ range
and was derived from laboratory
spectra. Table~\ref{tbl.fe7.wavelengths} gives the \citet{ekberg81}
wavelengths and the measured EIS velocities from the present spectrum
for those \ion{Fe}{vii} lines from Ekberg's list that are clearly
identified in the EIS spectra and that are either unblended or provide
the dominant contribution to blended lines. Table~6 in Paper~I
presented velocities from the cool ions presented in that paper and it
was noted all ions between \ion{O}{iv} ($\log\,T=5.21$) and \ion{Mg}{vii}
($\log\,T=5.76$) have velocities of around 40~\kms. Taking the average
velocity of those lines identified to be unblended and without
anomalies in Table~6 of Paper~I, we derive a value of 40.4~\kms\
(using the velocities in the $v_{\rm ref}$ column of this table), with
a standard deviation of 4.5~\kms. \ion{Fe}{vii} belongs to this group
of cool ions, even with the uncertainty in the ion fraction discussed
in Sect.~\ref{sect.ion-fraction}, and so we expect the ion's line
velocities to be consistent with the other cool ions. Therefore in
Table~\ref{tbl.fe7.wavelengths} we indicate those emission lines for
which the measured velocity is not consistent with the cool ion
velocity of 40.4~\kms. We find 18 of the 25 lines show good agreement,
giving confidence in the \citet{ekberg81} measurements. The anomalous
lines will be discussed in the following sections.

Since a number of new line identifications have been performed for
\ion{Fe}{vii} in this work, it is necessary to convert the measured
wavelengths to 
rest wavelengths in order to derive new experimental energies for the
upper emitting levels. For this we subtract the average cool ion velocity
of 40.4~\kms\ from the measured wavelengths.

\begin{deluxetable}{ll}
\tablecaption{Measured velocities for \ion{Fe}{vii} lines.\label{tbl.fe7.wavelengths}}
\tablehead{Reference & \\
wavelength\tablenotemark{a} &Velocity\tablenotemark{b,c} \\
(\AA) & (\kms) }
\tablewidth{0pt}
\startdata
176.744 & $    30.5 \pm  10.9$ \\
182.071 & $    46.1 \pm  16.8$ \\
182.740 & $    24.6 \pm  14.7$\vtl \\
183.539 & $    44.1 \pm  11.1$ \\
183.825 & $    39.1 \pm   8.6$ \\
184.752\tablenotemark{d} & $    40.6 \pm   9.1$ \\
184.886 & $    58.4 \pm   9.8$\vtl \\
185.547 & $    43.6 \pm   8.5$ \\
186.657 & $    56.2 \pm  12.9$\vtl \\
187.235\tablenotemark{e} & $    44.8 \pm   8.3$ \\
187.692 & $    35.1 \pm   8.6$ \\
188.396\tablenotemark{f} & $    44.5 \pm   9.5$ \\
188.576 & $    42.9 \pm   9.0$ \\
189.450 & $    49.1 \pm   8.1$\vtl \\
195.391 & $    36.8 \pm   7.7$ \\
196.046 & $    42.8 \pm   7.9$ \\
196.423 & $    53.4 \pm   7.9$\vtl \\
201.855 & $    47.5 \pm   7.9$ \\
207.712 & $    44.7 \pm   8.1$ \\
208.167 & $    46.1 \pm  17.3$ \\
\noalign{\smallskip}
265.697 & $    46.3 \pm   6.6$ \\
289.678\tablenotemark{g} & $    50.7 \pm   6.8$\vtl \\
289.831\tablenotemark{g} & $    54.8 \pm   6.1$\btl \\
290.307\tablenotemark{g} & $    39.2 \pm   5.9$ \\
290.756\tablenotemark{g} & $    36.1 \pm   5.6$ \\

\enddata
\tablenotetext{a}{From \citet{ekberg81}.}
\tablenotetext{b}{Errors represent the EIS measurement errors combined with the $\pm 0.005$~\AA\ errors on the \citet{ekberg81} reference wavelengths.}
\tablenotetext{c}{A \vtl\ symbol beside the velocity measurement
  indicates that it is discrepant with the average cool line velocity
  of $+40.4$~\kms\  by $>1\sigma$; a \btl\ indicates a discrepancy of $>2\sigma$.}
\tablenotetext{d}{Blended with \ion{Ne}{v} \lam184.735 and
  \ion{Fe}{xi} \lam184.803.}
\tablenotetext{e}{Blended with \ion{Fe}{viii} \lam187.241.}
\tablenotetext{f}{Possibly blended with \ion{Mn}{ix} \lam188.48.}
\tablenotetext{g}{Wavelengths not directly measured by \citet{ekberg81}, but deduced from other wavelength measurements.}
\end{deluxetable}

\subsection{Lines in the EIS LW band}

The \ion{Fe}{vii} lines in the LW band consist of decays from two
$n=4$ terms, $3p^63d4s$ $^3D$ and $3p^63d4p$ $^1P$, together with
decays from the $3p^53d^3$ $^5S$, $^5D$ and $^5F$ terms. The latter
two each give rise to two groups of lines corresponding to decays to
the $^3P$ and $^3F$ terms in the ground $3p^63d^2$ configuration. Line ratios
formed from the $3p^53d^3$ lines are relatively insensitive
with regard density and temperature, and we compare with the strongest
line, \lam249.30, below. There is significant temperature
sensitivity when comparing lines from different configurations. We go
through the LW lines by multiplet, starting with the longest
wavelengths.

Four emission features are found between 289 and 291~\AA, at the very
end of the EIS wavelength range. These are principally due to the
\ion{Fe}{vii} 
$3d^2$ $^3F_J$ -- $3d4s$ $^3D_{J^\prime}$ transitions, which were
first identified by \citet{brown08} from EIS spectra. The strongest
line is a blend of the $J=4$ to $J^\prime=3$ and 3--2 transitions, and
also has the \ion{Si}{ix} \lam290.69 line in the short wavelength
wing. The feature was fit with two Gaussians forced to have the same
width, with the short wavelength Gaussian representing the
\ion{Si}{ix} line. The remaining three emission features are
unblended. Each of \lam289.68, \lam289.83 and \lam290.31 shows only
weak sensitivity to density and temperature relative to
\lam290.72+\lam290.76, and the comparison with theory is shown in
Table~\ref{tbl.fe7.insens}. \lam290.31 is discrepant with theory,
with the observed line being too strong, however both \lam289.68 and
\lam289.83 agree with theory within the error
bars. Table~\ref{tbl.fe7.wavelengths} shows that the measured
velocities of \lam289.68 and \lam289.83 are discrepant with the
standard cool line velocity.
%show discrepancies.
The reference wavelengths for the $3d^2$ $^3F_J$ -- $3d4s$
$^3D_{J^\prime}$ transitions given in Paper~I are obtained from the
energies of \citet{ekberg81}, who was able to obtain the $3d4s$
level energies indirectly by measuring $4s$--$4p$ transitions at
wavelengths 1000--1400~\AA\ and $3d$--$4p$ transitions at wavelengths
200--300~\AA. Since the EIS wavelengths are direct measurements of the
$3d$--$4s$ transitions then the velocity discrepancies for \lam289.68
and \lam289.83 may be due to uncertainties in the \citet{ekberg81}
$4s$ configuration energies.

The \ion{Fe}{vii} model predicts two lines arising from the
$3p^53d^3(^4P)$ $^5S_2$ level whose theoretical wavelengths are
270.40 and 271.20~\AA, with
the latter being stronger by around a factor 3. 
The spectra were
searched in this region for two lines with the same ratio and
wavelength separation, and whose images are consistent with a cool
line. The lines observed in the atlas at 271.068 and 271.729~\AA\ were found to match
and so we identify these with the $3p^53d^3(^4P)$ $^5S_2$ level for which we are
thus able to establish an experimental energy value for the first
time (Table~\ref{tbl.fe7}). The longer wavelength line was used to
revise the energy since the shorter wavelength line is blended (see below), and a velocity
shift of $-40.4$~\kms\ was applied to determine the rest wavelengths,
as described in the previous section.
The rest wavelengths of the two decays from the $^5S_2$ level are
then 271.067 and 271.692~\AA\ for the decays to the $^3P_1$ and
$^3P_2$ levels in the ground configuration, respectively. \lam271.69 seems to be
unblended and
comparing with the strongest
line from $3p^53d^3$ in the LW band, the \lam271.69/\lam249.30
ratio shows weak sensitivity to density and temperature and the
theoretical value agrees well with
observations (Table~\ref{tbl.fe7.insens}). The observed line at
271.068~\AA\ is a very broad feature, suggesting it is a blend of two
or more lines, and Paper~I showed that two \ion{O}{v} lines
contribute, although they can not fully account for the line's
intensity. The \ion{Fe}{vii} \lam271.07/\lam271.69 branching ratio is
0.31 and implies that \ion{Fe}{vii} contributes an intensity of
7.7~\ecss\ to the blend.

\begin{figure}[h]
\epsscale{0.8}
\plotone{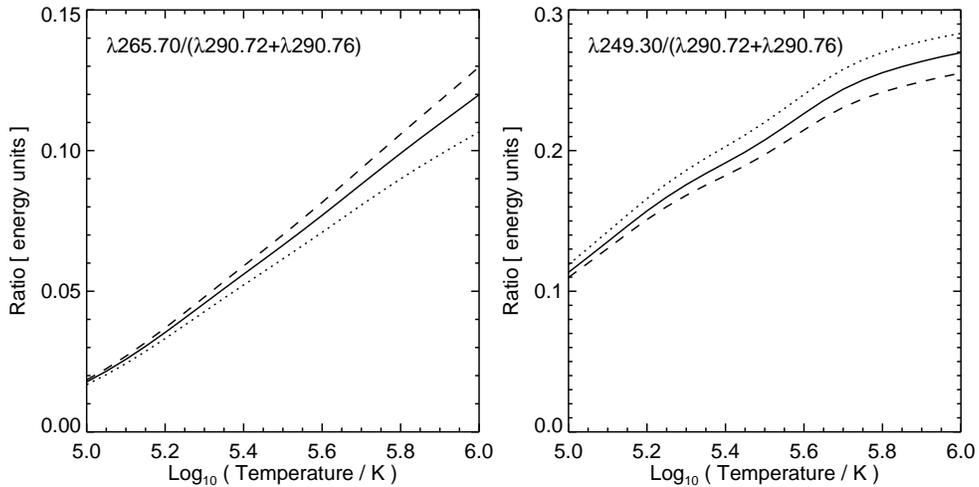}
\caption{Theoretical ratios formed from lines  found in the EIS LW
  band that arise from different configurations. The left panel shows
  \lam265.70/(\lam290.72+\lam290.76) and the 
right shows \lam249.30/(\lam290.72+\lam290.76). The ratios are plotted
as a function of temperature and the different lines correspond to
densities of $\log\,N_{\rm e}=$8.5 (dotted), 9.0 (solid) and 9.5 (dashed).}
\label{fig.lw-ratios}
\end{figure}

Around 265--267~\AA\ the \ion{Fe}{vii} model gives a line with known
wavelength at 265.70~\AA, and five lines with theoretical
wavelengths due to $3p^63d^2$ $^3P_J$ -- $3p^53d^3(^4F)$ $^5D_{J^\prime}$
transitions. The known line is the only decay from the $3p^63d4p$
configuration found in the EIS spectrum and
Table~\ref{tbl.fe7.wavelengths} shows the measured wavelength is
consistent with the laboratory wavelength of \citet{ekberg81}.
The 
\lam265.70/(\lam290.72+\lam290.76) ratio is found to be an excellent
temperature diagnostic (Fig.~\ref{fig.lw-ratios}), although it also shows some density
sensitivity. Taking the \ion{Mg}{vii} density of $\log\,N_{\rm
  e}$=9.15, we find a temperature of
$\log\,T=5.47^{+0.08}_{-0.07}$, which is close to the expected
temperature of formation of \ion{Fe}{vii}. Note that \lam265.70 
is blended with a high temperature \ion{Co}{xvi} line
\citep{brown08}, but this makes no contribution in the present
spectrum.

Studying images formed in the weak observed lines between 266 and 267.5~\AA,
the 267.29~\AA\ line is found to be close in temperature to
\ion{Fe}{vii}. The line is broad, suggesting it consists of more than
one transition, and we note that the 2--2, 2--1 and 2--3 transitions
of the $3p^63d^2$ $^3P_J$ -- $3p^53d^3(^4F)$ $^5D_{J^\prime}$ multiplet
are predicted to lie within 0.13~\AA\ of each other. Summing the theoretical
emissivities of each component and taking the ratio relative to 
\lam249.30 we find good agreement between theory and observation
(Table~\ref{tbl.fe7.insens}) suggesting that the observed line is
indeed a blend of three \ion{Fe}{vii} components.

Of the other observed emission lines between 266 and 266.7~\AA\, we believe
that a weak feature at around 266.42~\AA\ (not reported in Paper~I)
and the line measured at 266.623~\AA\ have some cool component based on
inspection of the images formed in the lines. The 1--1 and 1--0
components of the $3p^63d^2$ $^3P_J$ -- $3p^53d^3(^4F)$ $^5D_{J^\prime}$ multiplet are predicted to be
blended and to lie 0.7--0.8~\AA\ to the short wavelength side of the
earlier mentioned transitions and thus could be responsible for one or
both of these emission lines. \citet{brown08} reported a
\ion{Fe}{xvii} transition at 266.42~\AA\ but this will not be
significant in the present spectrum. 
The 266.62~\AA\ line was measured but not identified by
\citet{brown08}, and inspection of the line image suggests a line
formed at around 1--1.5~million~K in addition to the cool
component. Due to the weakness of the two 
lines and the unknown contribution of blending it is not possible to
make a definite identification of either of these observed lines with
specific \ion{Fe}{vii} 
transitions.

Our prescription for revising the energy levels for the 
$^5D_{J^\prime}$ term is to fit the broad line at 267.29~\AA\
with two Gaussians of the same width, giving components at 267.245 and
267.303~\AA\ with intensities 13.7 and 19.8~\ecss\ (Paper~I). The longer
wavelength component is assumed to be the 2--2 transition (which is
predicted to have the longest wavelength), yielding an
experimental energy for the $3p^53d^3(^4F)$ $^5D_{2}$ level of
395\,436~cm$^{-1}$. The shorter wavelength measured line is then
assumed to be a blend of the 2--1 and 2--3 components. Note that the
predicted ratio of the combined 2--1 and 2--3 transitions relative to
the 2--2 transition is 0.72 in very good agreement with the measured
ratio of 0.69. The $^5D_3$ energy value is derived using the
$3p^63d^2$ $^3F_3$ -- $3p^53d^3(^4F)$ $^5D_{3}$ transition later in
this section
which yields a predicted wavelength for $3p^63d^2$ $^3P_2$ --
$3p^53d^3(^4F)$ $^5D_{3}$ consistent with the measured 267.245~\AA\
line. We can then estimate an energy for the $^5D_1$ level by assuming
that the measured wavelength at 267.245~\AA\ corresponds to the
$3p^63d^2$ $^3P_2$ -- $3p^53d^3(^4F)$ $^5D_{1}$ transition. The
derived energy of 395\,517~cm$^{-1}$ is accurate to only around
$\pm$100~cm$^{-1}$ as the identified line is blended.

Five lines from the $3p^63d^2$ $^3P_J$ -- $3p^53d^3(^4F)$ $^5F_{J^\prime}$
multiplet are potentially observable in the EIS spectrum but
experimental wavelengths are not available. They are predicted to lie in
the region 257--259~\AA\ and the strongest line belongs to the $J=2$
to $J^\prime =3$ transition. Image inspection for two lines at 259.226 and
260.707~\AA\ reveals they are cool lines, close in temperature to
\ion{Fe}{vii} and so are possible candidates. The 259.226~\AA\ line has
contributions from \ion{Cr}{vii} and \ion{Al}{vii} although these can
not fully account for the strength of the observed line
(Paper~I). Based on the strength of the \ion{Fe}{vii} \lam249.30 line
the $3p^63d^2$ $^3P_2$ -- $3p^53d^3(^4F)$ $^5F_{3}$ is too strong to
be consistent with the remaining intensity of the 259.226~\AA\ line,
and so we identify the observed 260.707~\AA\ line with this
transition. The new experimental energy for $3p^53d^3(^4F)$ $^5F_{3}$
is given in Table~\ref{tbl.fe7} and the rest wavelength of the
transition is 260.672~\AA. 
The
\lam260.67/\lam249.30 ratio is predicted to be weakly sensitive to density
and temperature and the theoretical ratio is given in
Table~\ref{tbl.fe7.insens}, however the observed ratio is around
25~\%\ lower. No other nearby, unidentified emission lines provide a
better match in terms of intensity and so we are confident that the
2--3 transition does correspond to the observed 260.71~\AA\ line.

The next strongest line from the
multiplet is the 1--2 transition which is predicted to be a factor
0.59 weaker and 0.64~\AA\ shorter in wavelength. Two candidates in the
spectrum are the lines at 259.99 and 260.29~\AA. Inspecting images of
both lines shows they have a cool component in addition to a hotter
component. The 259.99~\AA\ line is a known \ion{Fe}{xii} transition
and this is confirmed by the line image, while the 260.29~\AA\ line appears
to be formed around $\log\,T=6.0$. Due to the uncertainty in the
identification of the 1--2 transition, we are not able to offer an
experimental energy for the $(^4F)$ $^5F_2$ level. 

A group of four $3p^63d^2$ $^3F_J$ -- $3p^53d^3(^4F)$ $^5D_{J^\prime}$ (${J^\prime}$=2,3,4)
transitions are predicted between 250 and 253~\AA\ and some guidance
as to the location of these lines is given by the earlier
identification of the $3p^63d^2$ $^3P_2$ -- $3p^53d^3(^4F)$ $^5D_{2}$
transition. Firstly we note that the strongest line of the
$^3F_J$--$^5D_{J^\prime}$ multiplet is predicted to be the 4--4
transition.  The difference between the observed energy and theoretical
energy for the $^5D_2$ level derived previously is 3195~cm$^{-1}$. If
we assume the $^5D_4$ level energy is also discrepant by this amount
then the predicted wavelength of the 4--4 transition becomes 253.89~\AA. The
next strongest line from the multiplet is the 3--3 transition which,
again using the energy correction from the $^5D_2$ level, is expected 
at 253.45~\AA. This line is predicted to be a factor 0.46
weaker than the transition from the $^5D_4$ level. There is a group of
five lines close to these predicted wavelengths, the strongest being an
\ion{Fe}{viii} line at 253.98~\AA\ (Sect.~\ref{sect.fe8}). The best
matches with the 
\ion{Fe}{vii} lines are the two lines measured at 253.56 and
254.09~\AA. Images formed in both lines are consistent with other
\ion{Fe}{vii} images, and the measured ratio is close to the predicted
ratio (Table~\ref{tbl.fe7.insens}). In addition the
\lam254.06/\lam249.30 ratio is in agreement with theory. The new
experimental energies for the $^5D_3$ and $^5D_4$ levels are given in
Table~\ref{tbl.fe7}. With four of the five levels of the $^5D$
multiplet now assigned experimental energies, a revised energy for the
remaining level, $^5D_0$, can be estimated. The average difference
between the experimental and theoretical energies for $^5D_{1-4}$ is
$-3269$~cm$^{-1}$ and subtracting this from the $^5D_0$ theoretical
energy gives the value in Table~\ref{tbl.fe7}. We estimate this value
is accurate to around $\pm$~200~cm$^{-1}$.

Two lines of the $3p^63d^2$ $^3F_J$ -- $3p^53d^3(^4F)$ $^5F_{J^\prime}$
multiplet are predicted to lie close to the short wavelength edge of the
EIS LW band: the 4--4 transition at 246.17~\AA, and the 4--5
transition at 246.69~\AA. As for the previously discussed multiplet,
these wavelengths can be revised based on the $3p^63d^2$ $^3P_2$ --
$3p^53d^3(^4F)$ $^5F_3$ transition identified earlier at 260.678~\AA. The
difference between the experimental and theoretical energies for
$^5F_3$ that this
identification implied is 4413~cm$^{-1}$. Adjusting the theoretical
energies for the $^5F_4$ and $^5F_5$ levels by the same amount leads
to new predicted wavelengths of 248.55 and 249.08~\AA. The atomic
model predicts that the short wavelength line is a factor 0.58 times
weaker than the long wavelength line. Based on this information, we
can identify the \ion{Fe}{vii} lines with two lines at 248.67 and
249.33~\AA\ whose ratio is $0.49\pm 0.07$ and for which the images are
consistent with other \ion{Fe}{vii} lines. The new experimental
energies for the upper levels of the two transitions are given in
Table~\ref{tbl.fe7}, and the rest wavelengths are 248.635 and 249.295~\AA.
These two identifications mean that three of five levels of the $^5F$
multiplet now have experimental energies, and these can be used to
estimate improved energies for the remaining $^5F_1$ and $^5F_2$
levels. The average difference between theoretical and observed
energies for $^5F_{3-5}$ is $-4575$~cm$^{-1}$, and applying this to
the $^5F_{1,2}$ theoretical energies yields the energies listed in
Table~\ref{tbl.fe7}. These values should be accurate to around $\pm
200$~cm$^{-1}$. 

\lam249.30 is the strongest of the EIS \ion{Fe}{vii}
emission lines in the LW band arising from the $3p^53d^3$
configuration and the above paragraphs demonstrated that the ratios
amongst the $3p^53d^3$ lines are in good agreement with theory -- see
also Table~\ref{tbl.fe7.insens}. We now consider ratios of \lam249.30
against the strongest lines from the $n=4$ multiplets. 

\lam249.30/(\lam290.72+\lam290.76) is temperature sensitive while
showing relatively weak density sensitivity
(Fig.~\ref{fig.lw-ratios}). However, the observed ratio of 
$0.39\pm 0.04$ is outside the range of variability of the theoretical
ratio, being too high compared to theory by a factor of around
two. \lam249.30/\lam265.70 shows both temperature and density
sensitivity, but if we calculate the theoretical ratio for a density of
$\log\,N_{\rm e}=9.15$ and temperature of $\log\,T=5.6$ we get a value
of 2.83, which compares with the observed ratio of $6.07\pm 0.082$. It
thus appears that the atomic data from \citet{witthoeft08}
over-predicts the strength of the lines from the $n=4$ configurations
compared to the $3p^53d^3$ configuration, based on the EIS measurements.

This completes our survey  of the \ion{Fe}{vii} lines in the EIS LW
waveband based on the new atomic model. We finish by noting that
\citet{ekberg81} listed eight transitions 
arising from levels in the $3p^63d4p$ configuration 
that lie in the LW band. Only one of these, the $3p^63d^2$ $^1S_0$ --
$3p^63d4p$ $^1P_1$ transition at 265.70~\AA, was discussed above as the
predicted intensities of 
the remaining transitions are all too low to be measured by EIS. For
example, 
the strongest of the seven remaining transitions, $3p^63d^2$ $^3P_0$ --
$3p^63d4p$ $^3D_1$  at 246.86~\AA, is predicted to be 0.47 times
weaker than \lam265.70 by the atomic model. However, the instrument
effective area is lower by further factor of 0.26 at the shorter
wavelength, making it too weak to be detectable in the present spectrum.

\begin{deluxetable}{ll}
\tablecaption{New level energies for \ion{Fe}{vii}.\label{tbl.fe7}}
\tablehead{& Energy \\
Level  & (cm$^{-1}$)}
\tablewidth{0pt}
\startdata
$3p^53d^3(^4P)$ $^5S_{2}$  & 389\,342 \\
$3p^53d^3(^4F)$ $^5D_{0}$  & 395\,459\tablenotemark{a} \\
$3p^53d^3(^4F)$ $^5D_{1}$  & 395\,342\tablenotemark{a} \\
$3p^53d^3(^4F)$ $^5D_{2}$  & 395\,436 \\
$3p^53d^3(^4F)$ $^5D_{3}$  & 395\,496 \\
$3p^53d^3(^4F)$ $^5D_{4}$  & 395\,954 \\
$3p^53d^3(^4F)$ $^5F_{1}$  & 404\,507\tablenotemark{a} \\
$3p^53d^3(^4F)$ $^5F_{2}$  & 404\,761\tablenotemark{a} \\
$3p^53d^3(^4F)$ $^5F_{3}$  & 404\,893 \\
$3p^53d^3(^4F)$ $^5F_{4}$  & 404\,518 \\
$3p^53d^3(^4F)$ $^5F_{5}$  & 404\,452 \\
\noalign{\medskip}
$3p^53d^3(a ^2D)$ $^3F_{4}$ & 510\,158 \\
$3p^53d^3(a ^2D)$ $^3F_{3}$ & 513\,537\tablenotemark{a} \\
$3p^53d^3(a ^2D)$ $^3F_{2}$ & 516\,029\tablenotemark{a} \\
$3p^53d^3(^2H)$ $^3G_{4}$   & 512\,601 \\ 
$3p^53d^3(^2H)$ $^1H_{5}$   & 538\,566 \\
$3p^53d^3(^2F)$ $^3D_{1}$   & 546\,454\tablenotemark{a} \\
$3p^53d^3(^4P)$ $^3P_{0}$   & 559\,991\tablenotemark{a} \\
\enddata
\tablenotetext{a}{Energy derived from theoretical level splittings.}
\end{deluxetable}

\subsection{Lines in the EIS SW band}

All the \ion{Fe}{vii} lines expected in the EIS SW band are emitted
from the $3p^53d^3$ configuration and \citet{ekberg81} provided a
large number of line identifications based on laboratory
spectra. These identifications were determined by searching for
emission lines whose spacing indicated that they represented decays
from a single upper level to different lower levels in the ground
configuration -- the lower level energies being well known, and thus
the line separations are accurately predicted. \citet{ekberg81} also
used theoretical calculations of level energies and radiative decay
rates to identify the upper levels.  
We first make general comments about comparisons between the
\citet{ekberg81} line identifications and the \citet{witthoeft08} atomic
model.

The $3p^53d^3$ configuration consists of 110 fine structure levels
from 48 spectroscopic terms. There are many duplicate terms (for
example there are six $^3D$ terms) and so it is necessary to
differentiate them by specifying the parent terms of the $3d^3$
sub-shell. \citet{witthoeft08} did not specify the parent terms, and
so they have been derived separately using the AUTOSTRUCTURE atomic
code by one of
the present authors (P.R. Young). By matching with the level ordering of
the $3p^53d^3$ configuration levels of \citet{ekberg81} we confirm all
of the parent terms listed by \citet{ekberg81}, except for his
$(^2F)^1D_2$ level, which we find to be $(b ^2D)^1D_2$. 
For the levels identified by \citet{ekberg81}, the \citet{witthoeft08}
theoretical energy values are between 13\,000 and 23\,000~cm$^{-1}$ larger.

When comparing theoretical structure calculations with previously
identified energy levels, there is a risk of mismatches due to level
mixing. E.g., if $^1F_3$ and $^3D_3$ levels are strongly mixed, then
the level names become arbitrary and so one author may assign the name
``$^1F_3$'' to a level, while another author might assign
``$^3D_3$''. To ensure that the same level identifications are used,
it is necessary to study the strengths of the transitions predicted to
arise from the levels. To check this in the present case we have
gone through each level identified by \citet{ekberg81} and compared
his measured intensities for each line emitted by the level, and then
compared with the predictions from the new atomic  model. For all but one
level, the strongest line predicted by the model agrees with the
strongest line measured by \citet{ekberg81}. The one exception is the
$3p^53d^3(^2F)$ $^3D_2$ level for which the model gives the decay to $^3P_1$ to
be the strongest, whereas \citet{ekberg81} finds the decay to $^1D_2$
to be the strongest. However, the differences are not large and no
other nearby level in the model is consistent with the
\citet{ekberg81} intensities. We are thus confident that all the
\citet{ekberg81} levels are correctly matched with levels of the same
name in the \citet{witthoeft08} model.

In the following paragraphs we will systematically go through each of
the eleven spectroscopic terms in the \ion{Fe}{vii} $3p^53d^3$ configuration
that  give rise to the EIS SW lines, starting with
the longest wavelength lines. In order to check the consistency of the
line intensities we will make three types of comparison: (i) ratios of
lines emitted from the same upper level (branching ratios); (ii)
ratios relative to the strongest line from a multiplet; and (iii)
ratios relative to the strongest line in the EIS SW band, the
$3p^63d^2$ $^3F_4$ -- $3p^53d^3(^2H)$ $^3G_5$ transition, \lam195.39.
Since all of the \ion{Fe}{vii}
transitions in the EIS 
SW band are due to $3p^63d^2$ 
-- $3p^53d^3$ transitions, we will not refer to configurations in the
notation below. Thus, e.g., $^3F_4$--$(^2H)^3G_5$ refers to the
$3p^63d^2$ $^3F_4$ -- $3p^53d^3(^2H)$ $^3G_5$ transition.

Working through the \ion{Fe}{vii} lines from the longest wavelengths
in the SW band through to the shortest, we begin with the two lines
from the $(^2F)^3G_3$ level:
% are expected to be seen in the 
% EIS spectra
the decay to $^3F_2$ at 207.71~\AA\ and the decay to
$^3F_3$ at 208.17~\AA. The former is the stronger and the wavelength
is consistent with the \citet{ekberg81} laboratory measurement
(Table~\ref{tbl.fe7.wavelengths}). 
Comparing with the strongest 
\ion{Fe}{vii} line seen in the EIS spectrum, \lam195.39, the
\lam207.71/\lam195.39 ratio is found to be relatively insensitive to density and
temperature and the measured line ratio is around 25~\%\ larger than predictions
(Table~\ref{tbl.fe7.insens}). The \lam208.17 line can also
be identified in the spectrum and, although very weak, the observed
intensity is consistent with the \lam207.71 line
(Table~\ref{tbl.fe7.branch}), while the derived velocity is also
consistent with other cool lines (Table~\ref{tbl.fe7.wavelengths}).

The strongest line from the $(^2F)^1G_4$ level is the decay to $^3F_3$
at 201.86~\AA. This line is observed in the EIS spectrum partly in the
short wavelength wing of the stronger \ion{Fe}{xiii} \lam202.04
line, and the wavelength is consistent with the \citet{ekberg81}
laboratory measurement (Table~\ref{tbl.fe7.wavelengths}).
The strength of \lam201.86 is compared with the strong
\lam195.39 \ion{Fe}{vii} line in Table~\ref{tbl.fe7.insens}, however
the observed line is stronger than predictions by a factor 2.
A further line from $(^2F)^1G_4$ is the decay to $^3F_4$ at
202.38~\AA, which is predicted to be a factor 5.0 less than
\lam201.86. In the EIS spectrum at this wavelength there is a feature
that can be fit with two Gaussians giving the two lines at 202.344 and
202.420~\AA\ in the line list table (Paper~I). Neither of these wavelengths is
consistent with the \ion{Fe}{vii} line. The stronger line at
202.42~\AA\ was identified as a blend of \ion{Fe}{xi} and
\ion{Fe}{xiii} by \citet{brown08}. However the \ion{Fe}{xiii}
identification is incorrect as the $3s^23p^2$ $^3P_1$ -- $3s^23p3d$
$^3P_0$ transition actually occurs at 203.16~\AA. An image formed in
the 202.42~\AA\ line is clearly consistent with \ion{Fe}{xi} and there
is a suggestion that a cool line contributes at the footpoint regions,
but this is not clear. The \ion{Fe}{vii} \lam202.38/\lam201.86
branching ratio implies that \ion{Fe}{vii} contributes an intensity of
10~\ecss\ to the measured line at 202.42~\AA.

The $^3F_J$ -- $(^2H)^3G_{J^\prime}$ multiplet is the most important
group of \ion{Fe}{vii} lines observed by EIS as they lie very close to
the peak sensitivity of the instrument, making them the strongest
\ion{Fe}{vii} lines. However there are problems reconciling the Ekberg
line identifications with the atomic model and the measured EIS line
intensities. The situation is complicated by the fact that
the  model predicts further lines from the $^3F$--$(a ^2D)^3F$
and $^1G$--$(^2H)^1H$ multiplets to lie close to the $^3F_J$ --
$(^2H)^3G_{J^\prime}$ multiplet.

The strongest lines from the $^3F_J$ -- $(^2H)^3G_{J^\prime}$ multiplet
are given by Ekberg at 195.39~\AA\ (4--5), 196.42~\AA\ (3--4), and
196.05~\AA\ (2--3). The atomic model predicts the relative strengths
to be 1.0:0.68:0.35. Lines can be seen corresponding to each of the
Ekberg lines in the EIS spectra, however their ratios are
1.0:0.20:0.36. The $^3F_3$ -- $(^2H)^3G_4$ line is thus clearly
discrepant with theory by a factor 3.

In the vicinity of these three lines in the EIS spectra we find two
unidentified lines at 195.51 and 196.24~\AA\ whose images are
consistent with \ion{Fe}{vii}. The measured ratios of these lines
relative to 195.40 are 0.52 and 0.51 and are thus more consistent with
the predicted intensity of the $^3F_3$ -- $(^2H)^3G_4$ transition. The
theoretical model of \ion{Fe}{vii} predicts that the $^3F_3$ --
$(^2H)^3G_4$ transition lies between the wavelengths of $^3F_4$ --
$(^2H)^3G_5$ and $^3F_2$ -- $(^2H)^3G_3$. We thus believe that the
observed line at 195.51~\AA\ is actually the $^3F_3$ -- $(^2H)^3G_4$
transition, and we use the measured wavelength to derive a new
experimental energy for the $(^2H)^3G_4$ level which is given in
Table~\ref{tbl.fe7}. The rest wavelength for the $^3F_3$ --
$(^2H)^3G_4$ transition is then 195.480~\AA. We stress that there is
some 
uncertainty over this identification on account of the fact that the
\lam195.48/\lam195.39 ratio is discrepant with the prediction from the
atomic model as shown in Table~\ref{tbl.fe7.insens} with
\lam195.48 apparently too weak in observations by around 25~\%,
however no better solution for the identification of this strong
transition can be found.

To explain the decays to $^3F_3$ and $^3F_4$ that Ekberg identified
with $(^2H)^3G_4$, we find that they are consistent with transitions
from $(a ^2D)^3F_4$. The theoretical data of \citet{witthoeft08} place
$(a ^2D)^3F_4$ very close in energy to the $(^2H)^3G$ levels, implying
the decays to the $^3F$ ground term will be close in wavelength. In
addition the predicted $^3F_4$ -- $(a ^2D)^3F_4$ to $^3F_3$ -- $(a
^2D)^3F_4$ intensity ratio is 0.44, consistent with Ekberg's measured
ratio of 0.47. (The predicted $^3F_4$ -- $(^2H)^3G_4$ to $^3F_3$ --
$(^2H)^3G_4$ intensity ratio is 0.04 and so is inconsistent with
Ekberg's measurements.) By making the identification of the $(a
^2D)^3F_4$ level with Ekberg's transitions we transfer Ekberg's
experimental energy for $(^2H)^3G_4$ to $(a ^2D)^3F_4$, and this
energy is given in Table~\ref{tbl.fe7}. No experimental energies were
available for the $(a ^2D)^3F_2$ and $(a ^2D)^3F_3$ levels and so we
have used the difference between the experimental and theoretical
energies for $(a ^2D)^3F_4$ to adjust the theoretical energies for
these two levels, and the new values are given in
Table~\ref{tbl.fe7}. With regard the EIS spectra, the \lam196.42 line
that we identify with $^3F_3$ -- $(a ^2D)^3F_4$ is partly blended with
\ion{Fe}{xiii} \lam196.52 \citep[note the revised wavelength of this
transition suggested by][]{young08} as well as another \ion{Fe}{vii}
line, the $^3F_3$ -- $(^2H)^3G_3$ transition at 196.45~\AA. Inspection
of the observed feature shows an emission line with an extended
shoulder on the long wavelength side, corresponding to dominant
\ion{Fe}{vii} emission and a weak \ion{Fe}{xiii} component. It can be
fit with two Gaussians forced to have the same width
(Table~4 of Paper~I) and the \ion{Fe}{xiii} component intensity results
in the \ion{Fe}{xiii} \lam196.54/\lam202.04 density diagnostic
yielding a density of $\log\,N_{\rm e}=9.0$, which is reasonably
consistent  with the \lam200.02/\lam202.04 ratio of the same ion.
If we assume the remaining component of the observed feature comprises
the \lam196.42 and \lam196.45 transitions then we can estimate the
\lam196.45 contribution as 8.5~\ecss\ (24~\%) based on the
\lam196.45/\lam196.05 branching ratio which has a value of 0.106.
This leaves an estimated intensity of $26.6\pm
1.4$~\ecss\ for \lam196.42. However, this is a factor two stronger
than expected based on the theoretical \lam196.42/\lam195.39 ratio
(Table~\ref{tbl.fe7.insens}). 

The strongest transition from the $(a ^2D)^3F$ term is $^3F_2$ -- $(a
^2D)^3F_3$, and with  the theoretical $(a ^2D)^3F_3$ energy revised as
above the predicted wavelength becomes 194.728~\AA. The predicted
ratio relative to \lam195.39 is $0.152\pm 0.007$, while that relative
to \lam196.42 is $2.28\pm 0.48$ placing the expected intensity
somewhere between 34 and 61~\ecss. A possible
candidate is the line measured at 194.82~\AA\ in the present
spectrum. Inspecting the image formed from this line suggests that it
is mostly due to \ion{Fe}{ix} (see Sect.~\ref{sect.fe9}),
however by forming an image towards the short wavelength
side of the line, a cool component similar to \ion{Fe}{vii} can be
seen. The line at 194.82~\AA\ is broader than the nearby
\ion{Fe}{viii} \lam194.66 and \ion{Fe}{ix} \lam197.86 lines suggesting
that it is a blend of two lines, and the measured intensity of the line is
consistent with it partly consisting of the \ion{Fe}{vii} $^3F_2$ -- $(a
^2D)^3F_3$ transition. Despite this we do not make a definite
identification of the \ion{Fe}{vii} transition with the observed line
and so do not revise the $(a ^2D)^3F_3$ energy. Analysis of high resolution
laboratory spectra may possibly be able to confirm the identification.

Returning to the $^3F_J$ -- $(^2H)^3G_{J^\prime}$  multiplet, we note
that the image formed in the \lam195.39 line shows a contribution from
a coronal line which is consistent with \ion{Fe}{x}. \citet{bromage77}
identified the \ion{Fe}{x} $3s^23p^5$ $^2P_{3/2}$ -- $3s^23p^4(^1S)3d$
$^2D_{3/2}$ transition with a line at 195.399~\AA\ in laboratory spectra, however
\citet{keenan08} identified the transition with a line at 195.32~\AA\
in solar spectra. Based on several different spectra taken with EIS,
\citet{brown08} identify the \ion{Fe}{x} transition with the
laboratory wavelength, and find no line corresponding with the line
observed by 
\citet{keenan08}. For this work we have inspected a quiet Sun off-limb
spectrum (where \ion{Fe}{vii} is negligible) and confirm that the
\ion{Fe}{x} wavelength is 195.40~\AA. Note that the \ion{Fe}{x} model
in CHIANTI does not give an observed wavelength for the transition,
instead the theoretical wavelength of 195.316~\AA\ is used -- see
\citet{delzanna04}. We can use the CHIANTI model to estimate the
contribution of \ion{Fe}{x} to the \ion{Fe}{vii} \lam195.39 line:
\ion{Fe}{x} \lam195.40/\lam184.54 is found from CHIANTI to be density
sensitive, ranging from 0.025 at $\log\,N_{\rm e}=8$ to 0.057 at
$\log\,N_{\rm e}=12$. If we use the density derived from \ion{Mg}{vii}
of $\log\,N_{\rm e}=9.15$, then the predicted \ion{Fe}{x} ratio is
0.032. The measured \ion{Fe}{x} \lam184.54 intensity is 504.3~\ecss,
giving a predicted \ion{Fe}{x} \lam195.40 intensity of 16~\ecss\ --
only a 7~\%\ contribution to the measured line at 195.42~\AA.

The \ion{Fe}{vii} $^3F_3$ -- $(^2H)^3G_4$ transition, \lam195.48,
lies close to the \ion{Ni}{xv} $3s^23p^2$ $^1D_2$ -- $3s^23p3d$ $^1D_2$
transition at 195.52~\AA\ \citep{brown08}, but this line is negligible
in the present spectrum as it is formed at $\log\,T=6.4$.  The $^3F_2$
-- $(^2H)^3G_3$ transition, \lam196.05, lies in the long wavelength wing of the
stronger \ion{Fe}{viii} \lam195.97 line, and a two Gaussian fit is necessary to
separate the components. 
\citet{brown08} list a \ion{O}{iv} transition between
the \ion{Fe}{viii} and \ion{Fe}{vii} lines, but there is no evidence
for this line here.

\begin{figure}[h]
\epsscale{0.8}
\plotone{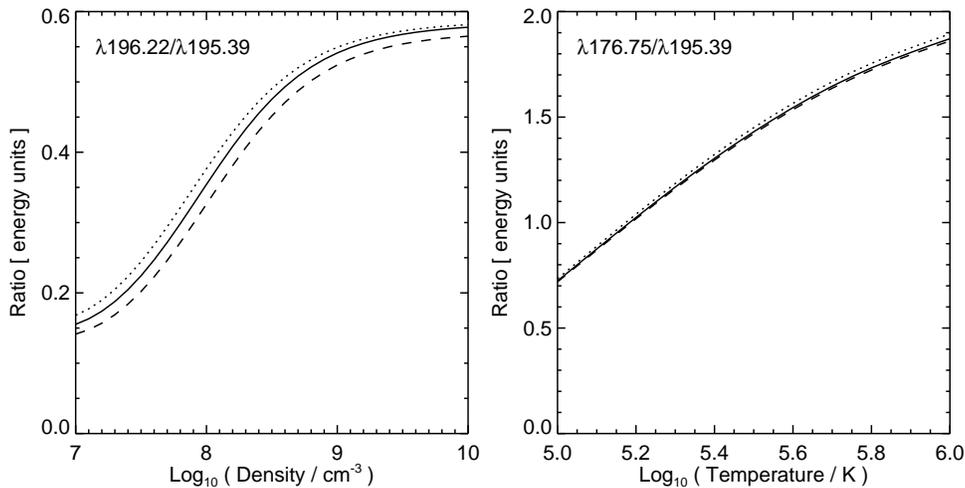}
\caption{Theoretical ratios formed from lines found in the EIS SW. The left panel shows
  the density sensitive \lam196.22/\lam195.39 ratio, with curves
  calculated at temperatures of $\log\,T=5.40$ (dotted line), 5.55 (solid line) and
  5.70 (dashed line). The right panel shows the temperature sensitive
  \lam176.75/\lam195.39 ratio, with curves calculated at densities of
  $\log\,N_{\rm e}=$8.5 (dotted), 9.0 (solid) and 9.5 (dashed).} 
\label{fig.sw-ratios}
\end{figure}

The $^1G_4$ -- $(^2H)^1H_5$ transition is predicted by the atomic
model to be 0.53 times the intensity of the $^3F_4$ -- $(^2H)^3G_5$
transition at 195.39~\AA, and the theoretical wavelength is 1.06~\AA\
longward of this transition. Therefore the observed line at
196.24~\AA\ is an excellent match. The only other line predicted to
arise from the $(^2H)^1H_5$ level in the EUV is the decay to
$^3F_4$ which occurs around 10~\AA\ shorter in wavelength compared to
the decay to $^1G_4$. The atomic model predicts this line to be 43 times weaker and so
it can not be observed by EIS.  The measured wavelength of the
196.24~\AA\ line is used to derive a new energy value for the
$(^2H)^1H_5$ level, and this is given in Table~\ref{tbl.fe7}. The rest
wavelength of the transition is then 196.217~\AA. Note that
\lam196.22/\lam195.39 is a good density diagnostic and
Fig.~\ref{fig.sw-ratios} shows the variation with density at three
temperatures. For the present spectrum, choosing the temperature of
$\log\,T=5.55$ we find a density of $\log\,N_{\rm e}=8.68\pm 0.08$,
which is significantly lower than the values from the \ion{Mg}{vii}
and \ion{Si}{vii} density diagnostics (Paper~I). To obtain the
\ion{Mg}{vii} density, the observed \lam196.22/\lam195.39 ratio would
have to be increased by only 10~\%\ which is a relatively small
discrepancy compared to other ratios discussed in this section.

The $(a ^2D)^1D_2$ level gives rise to only one line capable of being
seen by EIS: the decay to $^1D_2$ at wavelength 192.006~\AA. This line
is blended with \ion{Fe}{viii} \lam192.01 and a coronal line, which we
believe to be due to \ion{Fe}{xi}. The \ion{Fe}{vii}
\lam192.01/\lam195.39 ratio is weakly sensitive to density and
temperature with a value of $0.041\pm 0.007$, and so the \ion{Fe}{vii}
contribution to the 192.01~\AA\ line can be estimated as $9.2\pm
2.5$~\ecss\ (12~\%\ of the total line intensity),
using the measured \lam195.39 intensity. In Sect.~\ref{sect.fe8} the
\ion{Fe}{viii} contribution to the observed 192.02~\AA\ line is
estimated to be 54~\%.

\citet{ekberg81} identified lines from the $J=2$ and $J=3$ components
of the $(^2F)^3D_J$ multiplet, but not from the $J=1$ component. We
discuss first the identified lines. The strongest line predicted by
the atomic model is the $^3P_2$ -- $(^2F)^3D_3$ transition at
188.58~\AA\ which lies in the wing of the stronger \ion{Fe}{ix}
\lam188.50 line. A simultaneous 15 Gaussian fit was performed over the
range 187.71 to 189.13~\AA\ due to the lack of nearby continuum around
these lines.
The \ion{Fe}{vii} line
is found to be much weaker than expected: \lam188.58/\lam195.39 has a
theoretical value of $0.489\pm 0.068$, yet the observed ratio is
$0.253\pm 0.034$. The next strongest line from the $(^2F)^3D_3$ level
is the decay to $^1D_2$ at 187.24~\AA\ which is blended with
\ion{Fe}{viii} \lam187.24 and, as noted in Sect.~\ref{sect.fe8}, 
\ion{Fe}{vii} appears to contribute more than half of the measured
intensity. This actually makes it a little stronger than \lam188.58
whereas theory predicts it to be almost a factor four weaker. A further
line emitted from the $(^2F)^3D_3$ level is the decay to $^3F_4$ at
182.07~\AA. Although predicted to be only 0.13 the strength of 
\lam188.58, a line can be identified in the wing of \ion{Fe}{xi}
\lam182.16 that appears to be the \ion{Fe}{vii} transition. The line
fit parameters in the line list table of Paper~I are derived by forcing the
\ion{Fe}{vii} line to have the EIS instrumental width of 56~m\AA,
otherwise an unrealistically narrow line results. The
line velocity is consistent with the other
\ion{Fe}{vii} transitions (Table~\ref{tbl.fe7.wavelengths}), but the
\lam182.07/\lam188.58 ratio of 
$0.407\pm 0.164$ is significantly discrepant with theory. These
results suggest that the atomic data for the $(^2F)^3D_3$ level are in error.

The strongest line predicted from the $(^2F)^3D_2$ level is at
189.45~\AA\ and this can be identified in the EIS spectrum.
Comparing
with the strongest line from the $(^2F)^3D_3$ level, the
\lam189.45/\lam188.58 ratio shows weak sensitivity to density and
temperature but the observed ratio is almost a factor two stronger
than the predicted ratio (Table~\ref{tbl.fe7.insens}). This is consistent with the problems
discussed in the previous paragraph with regard the $(^2F)^3D_3$
level, and we note that \lam189.45/\lam195.39 is in better agreement
with theory (Table~\ref{tbl.fe7.insens}). The next strongest line
predicted from $(^2F)^3D_2$ is the decay to 
$^1D_2$ at 188.40~\AA\ which, however, is observed to be stronger than \lam189.45: the
observed \lam188.40/\lam189.45 branching ratio is almost a factor two larger than
the predicted value (Table~\ref{tbl.fe7.insens}). 
Table~\tblll\ in Paper~I lists a \ion{Mn}{ix} transition as a possible
blend to \lam188.40, contributing 30~\%\ to the observed line intensity based on the
DEM intensity prediction. As mentioned in Paper~I, the experimental
wavelength of the \ion{Mn}{ix} line is 
only accurate to $\pm0.05$~\AA\ and so it could actually be blending
with the nearby \ion{Fe}{ix} \lam188.50 line, but the large \ion{Fe}{vii}
branching ratio discrepancy suggests it is more likely to blend with
\ion{Fe}{vii} \lam188.40.
A decay to the level
$^3F_3$ gives a line at 182.74~\AA\ which is seen in the EIS spectra,
although the wavelength shows a small discrepancy compared to the
\citet{ekberg81} laboratory wavelength (Table~\ref{tbl.fe7.wavelengths}).
The \lam182.74/\lam189.45 branching ratio is consistent with
theory (Table~\ref{tbl.fe7.insens}), however. A further decay to $^3P_2$ is
predicted at 189.76~\AA, with the \lam189.76/\lam189.45 branching
ratio suggesting an intensity of around 9~\ecss. A line at this
wavelength can not be clearly seen in the spectrum, however.

\citet{ekberg81} did not report any lines from the $(^2F)^3D_1$ level,
however the atomic model predicts three potentially observable
lines. The \citet{witthoeft08} atomic data yield predicted wavelengths
for these lines, but more accurate predicted wavelengths can be
obtained by making use of the experimental energies of the
$(^2F)^3D_{2,3}$ levels. We first note that the \citet{witthoeft08}
theoretical energy for $(^2F)^3D_1$ is
568\,459~cm$^{-1}$, while the average difference between the
theoretical and observed energies for $(^2F)^3D_{2,3}$ is
$+22006$~cm$^{-1}$. We thus estimate a revised energy for $(^2F)^3D_1$
of 546\,454~cm$^{-1}$ which has an accuracy of around
$\pm$2000~cm$^{-1}$. The strongest line emitted by $(^2F)^3D_1$ is 
the decay to $^3P_0$ with a predicted wavelength of 189.97~\AA\ which
is accurate to around $\pm 0.7$~\AA. It is
insensitive relative to the strongest line from $(^2F)^3D_2$
(\lam189.45) with a
value $0.508\pm 0.040$, which makes it a good match for the line
observed at 189.36~\AA: the observed ratio being $0.574\pm
0.039$. Images formed in the 189.36~\AA\ line are also consistent with
\ion{Fe}{vii}. If this identification is correct then we also expect
two further lines at 182.436 and 189.499~\AA\ in the spectrum whose
ratios relative to the stronger line are 0.43 and 0.68,
respectively. There is a weak line seen at 182.430~\AA\ however the ratio
relative to the 189.36~\AA\ line is $0.708\pm 0.185$, and so higher
than theory predicts. If a line exists at 189.499~\AA\ it will partly
blend with the observed \ion{Fe}{vii} line at 189.481~\AA. The line
width of the latter is not anomalously broad and so the predicted line
at 189.499~\AA\ is either not there, or anomalously weak. Because of
these problems we do not identify the observed 189.359~\AA\ line with
the \ion{Fe}{vii} $^3P_0$ -- $(^2F)^3D_1$ transition.

The strongest transition predicted from the $(b ^2D)^1D_2$ level is the
decay to $^1D_2$ at 186.66~\AA, which places it in the wing of the
strong \ion{Fe}{viii} \lam186.60 line. Performing a two Gaussian fit
to \lam186.60 reveals a weak line in the long wavelength wing, however
the wavelength is longer than expected based on the reference
\ion{Fe}{viii} lines
(Table~\ref{tbl.fe7.wavelengths}). \citet{brown08} identified the line
at this wavelength as a \ion{Ni}{xiv}
transition, however the intensity predicted for this line using the
DEM is $< 1$~\ecss\ and so it can be ignored in the present case.
If we assume that \ion{Fe}{vii} accounts entirely for the measured
line, then the discrepant wavelength suggests that the two Gaussian
fit may not  be accurately measuring the weak line.
This is also suggested by a look at the  \lam186.66/\lam195.39 ratio,
which is weakly sensitive to temperature and density:
Table~\ref{tbl.fe7.insens} shows that \lam186.66 is weaker than
expected by more than a factor two.
Two
further lines are predicted to arise from the  $(b ^2D)^1D_2$ level:
the decays to the $^3P_{1,2}$ ground levels at wavelengths 187.69 and
187.99~\AA, respectively, and the branching ratios are 0.13 and 0.11.
Based on the measured intensity of
\lam186.66 the lines should be very weak, but a line is observed at
187.71~\AA\ that is  consistent with the expected position of
\lam187.69 (Table~\ref{tbl.fe7.wavelengths}) and for which the image
is consistent with \ion{Fe}{vii}. The observed intensity, however, is
much stronger than expected (Table~\ref{tbl.fe7.branch}), with
the measured line at
187.71~\AA\ almost half the strength of \lam186.66. A line is
observed at 187.972~\AA\ but we believe this is a \ion{Fe}{ix} transition
(Sect.~\ref{sect.fe9}). \ion{Fe}{vii} \lam187.99 will lie in the long
wavelength wing of the this line, but no significant feature
is found here implying the line is weak.

The $(^2G)^1F_3$ level gives rise to two lines at 185.55~\AA\ and
186.87~\AA, and the branching ratio \lam186.87/\lam185.55 is
0.218. Lines at both wavelengths are observed, but \lam186.87 is
blended with a \ion{Fe}{xii} feature that is itself a blend of two
lines at 186.85 and 186.89~\AA. Assuming \lam185.55 is unblended,
\ion{Fe}{vii} is predicted to contribute $14.4\pm 0.9$~\ecss\ (8~\%)
to the measured feature at 186.88~\AA.  \lam185.55/\lam195.39 is
relatively insensitive to density and temperature, but the observed
ratio is around 40~\%\ below the predicted value
(Table~\ref{tbl.fe7.insens}). 

\citet{ekberg81} identified emission lines from the $J=1$ and $J=2$
components of the $(^4P)^3P_J$ term, but for the $J=0$ component only one
line is predicted by the atomic model -- the decay to $^3P_1$ -- and
so Ekberg's method of identifying multiple lines from a single level
can not be applied. An improved estimate of the $^3P_1$ --
$(^4P)^3P_0$ wavelength can be made, however, by taking the average
energy difference between the observed and theoretical energies for
the $(^4P)^3P_{1,2}$ levels, and applying this to the theoretical
energy of $(^4P)^3P_0$. The resulting energy is given in
Table~\ref{tbl.fe7} and we estimate has an accuracy of around
$\pm$500cm$^{-1}$. The predicted wavelength for $^3P_1$ --
$(^4P)^3P_0$ is then 185.34~\AA\ with an accuracy of around $\pm
0.2$~\AA.  This line is insensitive relative to the 
strongest line from the $(^4P)^3P_J$ term,
\lam183.83, with a theoretical value of $0.250\pm 0.024$, and so should
have an intensity around 23~\ecss. No obvious candidate in the EIS
spectrum can be found, but possibly it is blended with the strong
\ion{Fe}{viii} \lam185.21 line.

The $(^4P)^3P_1$ level gives rise to three lines of comparable
strength at wavelengths 184.75, 184.89 and 185.18~\AA, corresponding
to decays to the $^3P_{0,1,2}$ levels in the ground
configuration. They are each predicted to be around one quarter of the
strength of the $^3P_2$ -- $(^4P)^3P_2$ transition at 183.83~\AA.
\lam185.18 is blended with the strong \ion{Fe}{viii} \lam185.21 line
and makes a $<$~2~\% contribution. Table~\tblll\ of Paper~I indicates
that \lam184.89 is blended with \ion{Ne}{vi}, although it was noted
that the measured wavelength is not consistent with the \ion{Ne}{vi}
wavelength. If the measured intensity is assumed to be entirely due to
\ion{Fe}{vii} then 
\lam184.89/\lam183.83 is in good
agreement with theory (Table~\ref{tbl.fe7.insens}). The measured
wavelength, however, shows a discrepancy of around 20~\kms\
(Table~\ref{tbl.fe7.wavelengths}). The image formed in the line is
consistent with the formation temperatures of both \ion{Ne}{vi} and
\ion{Fe}{vii}. 
Table~\tblll\ in Paper~I also indicates that \ion{Fe}{vii} \lam184.75
is blended with \ion{Fe}{xi} and \ion{Ne}{v}, and based on the
measured \ion{Fe}{vii} \lam183.83 line intensity \ion{Fe}{vii} would be expected
to contribute around 60~\%. However, the DEM predictions for
\ion{Ne}{v} and \ion{Fe}{xi} account for 80~\%\ of the measured
intensity. No other \ion{Ne}{v} line is found in the spectrum so an
independent check of the line's intensity is not possible (Paper~I),
while the \ion{Fe}{xi} line is strongly density dependent relative to other
\ion{Fe}{xi} lines such as \lam188.23 so the predicted contribution
depends critically on the density chosen for the DEM analysis. We note
that the measured wavelength of the 184.777~\AA\ line is consistent
with \ion{Fe}{vii} \lam184.75  (Table~\ref{tbl.fe7.wavelengths}).
The image formed
in the \lam184.75 line clearly reveals a blend with a coronal line with
temperature around that of \ion{Fe}{xi}, however a cool component can
also be seen. 

Two potentially observable lines are predicted from the $(^4P)^3P_2$
level: the decay to $^3P_1$ at 183.54~\AA\ and the decay to $^3P_2$ at
183.83~\AA. Lines at both wavelengths are found, but the observed 
\lam183.54/\lam183.83 ratio is a factor 2 lower than theory
(Table~\ref{tbl.fe7.insens}). The 
\lam183.83/\lam195.39 ratio is relatively insensitive to density and
temperature, and theory agrees well with observation
(Table~\ref{tbl.fe7.insens}). 

The shortest wavelength \ion{Fe}{vii} lines observed by EIS are the
$^3F_J$ -- $(^4F)^3F_{J^\prime}$ transitions, and the 4--4 transition
at 176.75~\AA\ is in fact the strongest of all the \ion{Fe}{vii} lines
predicted by the atomic model for typical coronal conditions. The low
effective area at this wavelength means the line is rather weak in the
EIS spectrum, but it appears to be unblended. The
\lam176.75/\lam195.39 ratio is weakly sensitive to density but does
show temperature sensitivity as shown in Fig.~\ref{fig.sw-ratios}. Using
the density of $\log\,N_{\rm e}=9.15$ derived from \ion{Mg}{vii}
\lam280.72/\lam278.39 (Paper 1) we find a
temperature of $\log\,T=5.07\pm 0.10$, much lower than the 
formation  temperature of the ion. To yield a temperature of
$\log\,T=5.6$ (which we believe to be the $T_{\rm max}$ of the ion)
would require an observed ratio of 1.55. Therefore the actual observed
ratio is around a factor two lower than expected.

Two additional lines from the $^3F_J$ -- $(^4F)^3F_{J^\prime}$
multiplet are predicted to be observed -- the 3--3 transition at
176.93~\AA\ and the 2--2 transition at 177.17~\AA\ -- but both are blended
with stronger lines from other species. \lam176.93 is blended 
with a stronger \ion{Fe}{ix} line (Sect.~\ref{sect.fe9}) and we estimate the \ion{Fe}{vii}
contribution to be $132.8\pm 28.1$~\ecss\ as the \lam176.75/\lam176.93
ratio has weak sensitivity to density and temperature with a
theoretical value of $0.712\pm 0.085$. We note that the measured
line is broad, consistent with a blend of two lines with slightly
different wavelengths.  \lam177.17 is blended with \ion{Fe}{x}
\lam177.24 and, using the \lam177.17/\lam176.75 theoretical ratio of
$0.505\pm 0.045$ we estimate a \ion{Fe}{vii} contribution of $94.2\pm
18.4$~\ecss. 

\begin{figure}[h]
\epsscale{0.8}
\plotone{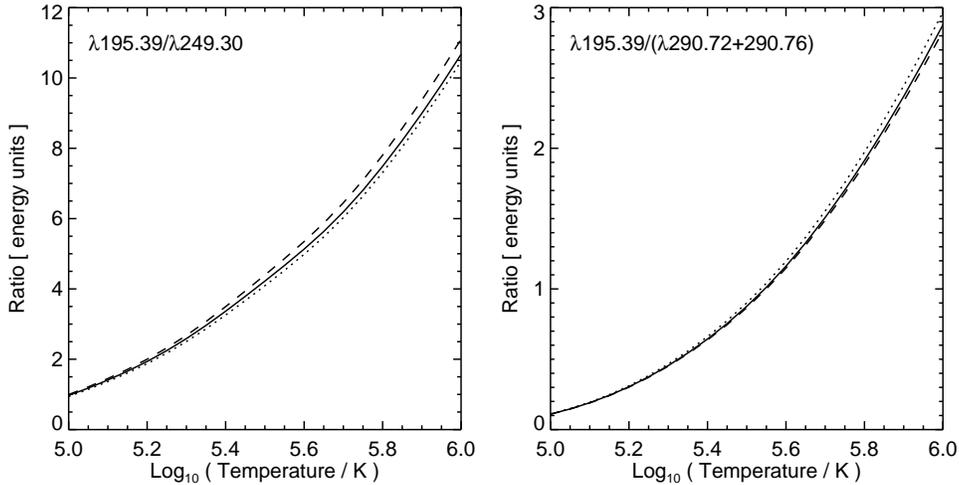}
\caption{Theoretical ratios formed from lines found in the two EIS
  wavelength bands. The left panel shows \lam195.39/\lam249.30 and the
right shows \lam195.39/(\lam290.72+\lam290.76). The ratios are plotted
as a function of temperature and the different lines correspond to
densities of $\log\,N_{\rm e}=$8.5 (dotted), 9.0 (solid) and 9.5 (dashed).}
\label{fig.lw-sw}
\end{figure}

Finally, we finish this section by comparing lines from the two EIS
wavelength bands. The strongest lines emitted by the $3p^53d^3$
configuration in the two bands are \lam195.39 and \lam249.30. Their
ratio is weakly sensitive to density, but strongly temperature
sensitive (Fig.~\ref{fig.lw-sw}) and the measured ratio gives a temperature of
$\log\,T=5.37^{+0.04}_{-0.02}$ which is close to the $T_{\rm max}$
value given by \citet{bryans09} but less than the temperature where we
believe \ion{Fe}{vii} is actually formed at. For the ratio to yield
higher temperatures, the measured \lam249.30 intensity would have to be weaker
than what is actually observed. Comparing \lam195.39 with the
\lam290.72+\lam290.76 self-blend that arises from the $3p^63d4s$
configuration, the ratio is again temperature sensitive, yielding a
temperature of $\log\,T=5.63\pm 0.02$ which is more consistent with
the apparent formation temperature of \ion{Fe}{vii}.

\begin{deluxetable}{lllrl}
\tablecaption{Fe VII branching ratios.\label{tbl.fe7.branch}}
\tablehead{Upper level &Ratio & Theory & Observation\tablenotemark{a}& }
\tablewidth{0pt}
\startdata
$3p^53d^3(^4P)$ $^3P_2$ &\lam183.54/\lam183.83  & 0.269  & $0.132\pm 0.032$ &\btl \\
$3p^53d^3(^4P)$ $^3P_1$ &\lam184.75\tablenotemark{b}/\lam184.89  & 1.07  & $1.80\pm 0.29$ & \\
$3p^53d^3({\rm b}^2D)$ $^1D_2$ &\lam187.69/\lam186.66 & 0.127 &$0.459
\pm 0.086$ &\btl \\
$3p^53d^3(^2F)$ $^3D_3$ &\lam182.07/\lam188.58 & 0.129 & $0.407\pm 0.164$ &\vtl\\
                       &\lam187.23/\lam188.58 & 0.247 & $1.095\pm 0.162$\tablenotemark{c} &\btl\\
$3p^53d^3(^2F)$ $^3D_2$ &\lam182.74/\lam189.45  & 0.172 & $0.202\pm 0.068$ &\\
                       &\lam188.40/\lam189.45  & 0.643 & $1.26\pm 0.15$ &\btl\\
$3p^53d^3(^2F)$ $^3G_3$ &\lam208.17/\lam207.71  & 0.141 & $0.109\pm 0.063$ &\\
\enddata\
\tablenotetext{a}{A \vtl\ symbol  indicates that the observed ratio is
  discrepant with theory by $>1\sigma$; a \btl\ indicates a discrepancy of $>2\sigma$.}
\tablenotetext{b}{Blended with \ion{Ne}{v} \lam184.735 and
  \ion{Fe}{xi} \lam184.803.}
\tablenotetext{c}{The \ion{Fe}{viii} contribution to \lam187.23 has
  been subtracted.}
\end{deluxetable}

\begin{deluxetable}{llll}
\tablecaption{Fe VII insensitive ratios.\label{tbl.fe7.insens}}
\tablehead{Term\tablenotemark{a} &Ratio & Theory & Observation\tablenotemark{b} }
\tablewidth{0pt}
\startdata
$3p^53d^3(^4P)$ $^3P$ &\lam184.89\tablenotemark{c}/\lam183.83  & $0.238\pm 0.007$ & $0.222\pm 0.034$ \\
 &\lam183.83/\lam195.39  & $0.437\pm 0.065$ & $0.404\pm 0.025$ \\
\noalign{\smallskip}
$3p^53d^3(^2G)$ $^1F$ &\lam185.55/\lam195.39  & $0.478\pm 0.078$ & $0.294\pm 0.018$ \btl \\
\noalign{\smallskip}
$3p^53d^3({\rm b}^2D)$ $^1D$ &\lam186.66/\lam195.39  & $0.458\pm 0.069$ & $0.195\pm 0.032$ \btl \\
\noalign{\smallskip}
$3p^53d^3(^2F)$ $^3D$ &\lam189.45/\lam188.58  & $0.542\pm 0.028$ & $0.935\pm 0.129$ \btl\\
 &\lam189.45/\lam195.39  & $0.265\pm 0.024$ & $0.236\pm 0.010$ \vtl \\
 &\lam188.58/\lam195.39  & $0.489\pm 0.068$ & $0.253\pm 0.034$ \btl \\
\noalign{\smallskip}
$3p^53d^3(^2H)$ $^3G$ &\lam195.48/\lam195.39  & $0.698\pm 0.057$ & $0.520\pm 0.013$ \btl \\
 &\lam196.05/\lam195.39  & $0.359\pm 0.026$ & $0.358\pm 0.013$ \\
\noalign{\smallskip}
$3p^53d^3({\rm a}^2D)$ $^3F$ &\lam196.42/\lam195.39  & $0.067\pm 0.015$ & $0.118\pm 0.006$\tablenotemark{d} \btl  \\
\noalign{\smallskip}
$3p^53d^3(^2F)$ $^1G$ &\lam201.86/\lam195.39  & $0.085\pm 0.017$ & $0.212\pm 0.013$ \btl \\
\noalign{\smallskip}
$3p^53d^3(^2F)$ $^3G$ &\lam207.71/\lam195.39  & $0.319\pm 0.034$ & $0.397\pm 0.036$ \vtl \\
\noalign{\medskip}
$3p^53d^3(^4F)$ $^5F$ &\lam248.64/\lam249.30  & $0.593\pm 0.06$ & $0.493\pm 0.067$ \\
 &\lam260.67/\lam249.30  & $0.373\pm 0.032$ & $0.281\pm 0.035$ \vtl\\
\noalign{\smallskip}
$3p^53d^3(^4F)$ $^5D$ &\lam253.52/\lam254.05  & $0.457\pm 0.016$ & $0.385\pm 0.065$ \\
 &\lam254.05/\lam249.30  & $0.800\pm 0.105$ & $0.702\pm 0.085$ \\
 & (\lam267.21+\lam267.22+\lam267.27)/\lam249.30 & $0.516\pm 0.050$ & $0.483\pm 0.046$ \\
 & (\lam267.21+\lam267.22+\lam267.27)/\lam254.06 & $0.645\pm 0.021$ &
 $0.688\pm 0.075$ \\ 
\noalign{\smallskip}
$3p^53d^3(^4P)$ $^5S$ &\lam271.69/\lam249.30  & $0.375\pm 0.053$ & $0.351\pm 0.036$ \\
\noalign{\smallskip}
$3p^63d4s$ $^3D$ &\lam289.68/(\lam290.72+\lam290.76) & $0.193\pm 0.001$ & $0.213\pm
0.032$ \\
 &\lam289.83/(\lam290.72+\lam290.76) & $0.191\pm 0.002$ & $0.205\pm
0.029$ \\
 &\lam290.31/(\lam290.72+\lam290.76) & $0.267\pm 0.005$ & $0.389\pm
0.041$ \btl\\
\enddata
\tablenotetext{a}{Ratios are grouped according to the spectroscopic
  term of the upper emitting level. Ratios are formed either between lines
  emitted from the same term, or one of these lines relative to a
  reference line (\lam195.39 for the EIS SW band, and \lam249.30 for
  the LW band).}
\tablenotetext{b}{A \vtl\ symbol indicates a $>1\sigma$ discrepancy
  between theory and observation, \btl\ indicates a $>2\sigma$
  discrepancy.}
\tablenotetext{c}{Possibly blended with \ion{Ne}{vi} \lam184.95.}
\tablenotetext{d}{The contribution of \ion{Fe}{vii} \lam196.45 has
  been subtracted (see text for details).}
\end{deluxetable}

\subsection{Summary of Fe\,VII results}\label{sect.fe7.summary}

The survey of the \ion{Fe}{vii} lines in the previous sections have
revealed a number of new line identifications in both the SW and LW
bands of EIS. The atomic model constructed from the
\citet{witthoeft08} data for CHIANTI  yields several line ratios that are
good diagnostics of temperature or density. Comparing theory with the
EIS observations shows many areas where good agreement is found, but
also a number of significant problems. To summarise briefly the main
points:

\begin{itemize}
\item The ratio of lines from the $3d4p$ and $3d4s$ configurations,
  \lam265.70/(\lam290.72+\lam290.76), is a temperature diagnostic.
\item Within the LW band (246--291~\AA) the atomic model
  over-predicts the strength of the lines from the $3d4p$ and
  $3d4s$ configurations compared to the lines from the $3p^53d^3$
  configuration by a factor of around two.
\item \lam195.39/(\lam290.72+\lam290.76) is a temperature
  diagnostic and the derived temperature is close to the expected
  value for \ion{Fe}{vii}.
\item Comparing strongest lines from the eleven $3p^53d^3$
  spectroscopic terms in the SW band:
\begin{itemize}
\item lines from $(^2F)^3G$, $(^2H)^1H$, $(^2F)^3D_2$ and $(^4P)^3P$
  are generally consistent with the \lam195.39 reference line;
\item lines from $(^2F)^3D_3$, $({\rm b}^2D)^1D$, $(^2G)^1F$ and
  $(^4F)^3F$ are too weak relative to \lam195.39 by around a factor
  two;
\item lines from $(^2F)^1G$ are too strong relative to \lam195.39 by
  around a factor two.
\end{itemize}
\item \lam196.22/\lam195.39 is identified as a density
  diagnostic over $\log\,N_e=7.0$ to 9.0, but the derived density here
  is lower than found from other ions formed at a similar temperature.
\item \lam176.75/\lam195.39 is a temperature diagnostic, but the EIS
  intensities yield a very low temperature, suggesting the atomic
  model over-predicts the strength of \lam176.75 relative to
  \lam195.39 by a factor of around two.
\end{itemize}

The good agreement between the measured EIS wavelengths and
the \citet{ekberg81} reference wavelengths gives confidence in the
\citet{ekberg81} transition identifications, however the large line
ratio discrepancies suggest problems with the current \ion{Fe}{vii}
atomic data. A new study of high resolution laboratory spectra
would be extremely valuable for investigating the problems further. In
particular it is important to 
confirm some of the new and suggested line identifications found in
the present work, for example the 195.48 and 196.21~\AA\ lines and the
lines between 248 and 272~\AA\ that arise from the $3p^53d^3$ configuration.

\section{Fe\,VIII}\label{sect.fe8}

The ground configuration of \ion{Fe}{viii} has only two energy levels,
$^2D_{3/2,5/2}$, and the EUV spectrum in the range 160 to 260~\AA\
consists principally of decays from the levels in the $3p^53d^2$
configuration to this ground term. Line identifications are known for
all lines between 160 and 200~\AA, but for 200--260~\AA\ there are a
large number of transitions that are unidentified. The EIS instrument
observes the wavelength ranges 170--212 and 246--292~\AA\ and so some
of these transitions can be found, but we first focus on the known
transitions.

The three terms $3p^53d^2(^3F)$ $^2F$, $3p^64d$ $^2P$ and
$3p^53d^2(^1S)$ $^2P$ give rise to nine lines between 185 and 198~\AA,
all of which are observed in the present spectrum. Reference
wavelengths for these lines are available from \citet{ramonas80}, and
Table~\ref{tbl.fe8.wavelengths} shows the velocities derived from the
EIS spectrum using these rest wavelengths. The error bars are derived
from the measured centroid uncertainties (Paper~I), the estimated
uncertainty of $\pm 0.002$~\AA\ in the EIS wavelength scale
\citep{brown07}, and the uncertainties in the measured wavelengths of
\citet{ramonas80} of $\pm 0.003$~\AA. The velocities of the lines
unaffected by blending are all less than the typical cool line
velocity of $+40.4$~\kms\ (Sect.~\ref{sect.ion-fraction}), with an average of $+33.0$~\kms. Paper~I
noted that ions formed above $\log\,T=5.8$ showed smaller velocity
shifts of around $+20$~\kms\ thus the \ion{Fe}{viii} velocities could
indicate that it is intermediate between the cool ion and hot ion
populations. 

\begin{deluxetable}{lll}
\tablecaption{\ion{Fe}{viii} velocities.\label{tbl.fe8.wavelengths}}
\tablehead{Wavelength & Velocity \\
(\AA) & (\kms) }
\tablewidth{0pt}
\startdata
 185.213 & $   30.7 \pm     5.8$ \\
 186.601 & $   36.9 \pm     5.8$ \\
 187.237\tablenotemark{a} & $   43.2 \pm     7.5$ \\
 192.004\tablenotemark{b} & $   34.4 \pm     7.3$ \\
 193.967 & $   32.5 \pm     6.4$ \\
 194.662 & $   27.7 \pm     5.6$ \\
 195.972 & $   32.1 \pm     5.5$ \\
 196.650\tablenotemark{c} & $   21.4 \pm     5.6$ \\
 197.362 & $   38.0 \pm     5.5$ \\

\enddata
\tablenotetext{a}{Blended with \ion{Fe}{vii} \lam187.235.}
\tablenotetext{b}{Blended with \ion{Fe}{vii} \lam192.006 and an
  unknown hotter line.}
\tablenotetext{c}{Blended with \ion{Fe}{xii} \lam196.640.}
\end{deluxetable}

\begin{figure}[h]
\plotone{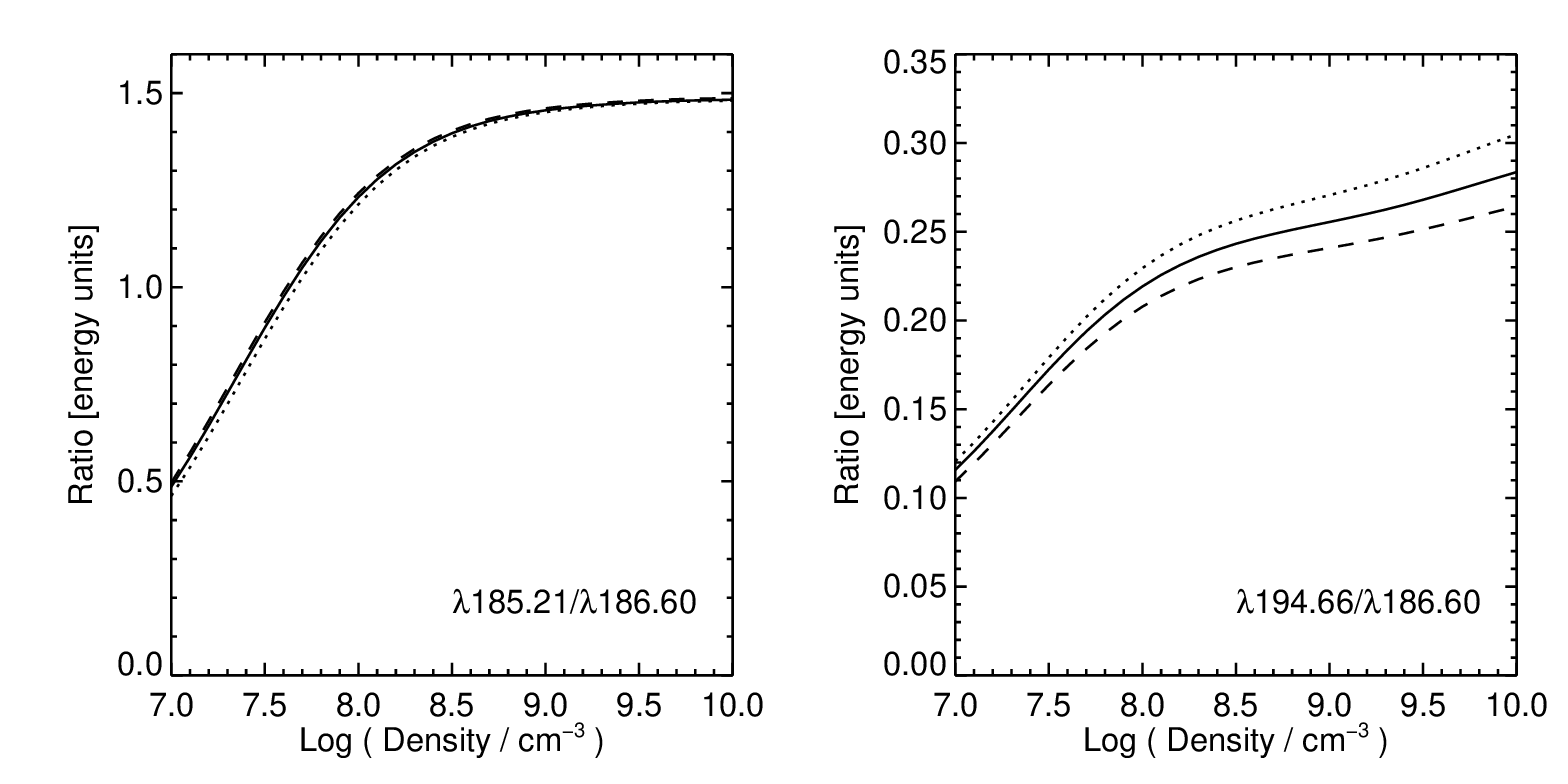}
\caption{\ion{Fe}{viii} emission line ratios showing density
  sensitivity. The dotted, solid and dashed lines have been calculated for temperatures
  $\log\,T=5.7, 5.8$ and 5.9, respectively. The left panel shows the
  \lam185.21/\lam186.60 ratio and the right panel the
  \lam194.66/\lam186.60 ratio.}
\label{fig.fe8.dens}
\end{figure}

\ion{Fe}{viii} line ratios formed from the nine lines between 185 and
198~\AA\ show very little density 
sensitivity above $10^{9}$~cm$^{-3}$ as the two ground levels are in
a quasi-Boltzmann equilibrium above this value. For this reason the
lines are little value as density diagnostics in typical coronal
conditions. Temperature sensitivity is also weak for these lines. 
Below $10^{9}$~cm$^{-3}$ significant density sensitivity sets in and
this is highlighted for 
the three strongest \ion{Fe}{viii} lines observed by EIS in
Fig.~\ref{fig.fe8.dens}. These ratios may be of value in coronal hole
or off-limb regions where the density is low, however, the present
spectrum reveals some anomalies. The observed \lam185.21/\lam186.60
ratio is found to be $1.28\pm 0.02$ which yields a density of
$\log\,N_{\rm e}=8.16\pm 0.05$, significantly below that found from
\ion{Mg}{vii} in Paper~I. There are high temperature blends for both
lines \citep{young07a}, but both are negligible in the present
spectrum. Since the observed ratio is only 14~\%\ below the high
density limit of the ratio we believe the discrepancy is most likely
due to atomic data uncertainties.
\lam194.66/\lam186.60 shows some temperature sensitivity in addition
to density sensitivity, however, the measured ratio is $0.402\pm
0.006$, significantly above the range of sensitivity of the
ratio. Since the two lines are both strong and unblended this, again,
we believe is due to inaccurate atomic data.

More generally we can investigate atomic data issues through studies
of insensitive line ratios. We divide the ratios into Groups 1, 2 and
3 which give an indication of the expected accuracy of the atomic
data. Group 1 consists of branching ratios, Group 2 consists of
insensitive ratios formed from lines belonging to the same multiplet,
while Group 3 consists of insensitive ratios formed from lines
belonging to different multiplets. Table~\ref{tbl.fe8.insens} presents
a comparison between the observed ratios and theoretical ratios
calculated with CHIANTI.
The atomic model for \ion{Fe}{viii} in CHIANTI
consists principally of the electron collision data and radiative
decay rates of \citet{griffin00}, supplemented with additional data
from \citet{czyzak66} and decay rates calculated by the CHIANTI team
\citep{dere01}. 

None of
the Group 1 measured ratios 
agrees with theory, although \lam193.97/\lam194.66 is within 11~\%\ of
the observed value. 
\lam187.24  is a known blend with \ion{Fe}{vii} \citep{brown08},
and thus we can use the \ion{Fe}{viii} ratio to estimate a
\ion{Fe}{vii} contribution of $62.2\pm 4.2$~\ecss, however
Sect.~\ref{sect.fe7} shows that this is inconsistent with the
\ion{Fe}{vii} atomic model. Note that the \ion{Fe}{viii} intensity predicted from the
DEM shown in Table~\tblll\ of Paper~I is significantly higher than
that predicted from the branching ratio. This is because the DEM
over-predicts the strength of the strong \lam186.60 line. \lam196.65 is a known
blend \citep{brown08, young08} with a \ion{Fe}{xii} transition. Using
the branching ratio to subtract the \ion{Fe}{viii} component leaves
an intensity of 50.2~\ecss\ for \ion{Fe}{xii}. \citet{young08} noted
that the \ion{Fe}{xii} \lam196.64/\lam186.88 ratio is relatively
insensitive to density, and we find a value of 0.27 after the
\ion{Fe}{viii} correction. In active regions where \ion{Fe}{xii} is
much stronger than \ion{Fe}{viii}, \citet{young08} found
\lam196.64/\lam186.88 ratios of between 0.24 and 0.32, consistent with
the value of 0.27 found here. This gives some confidence that the
\ion{Fe}{viii} \lam196.65/\lam197.36 branching ratio is consistent
with theory.

For  the Group 2 and 3 ratios the theoretical values are evaluated as
the 
averages of the ratios calculated over the density range
$\log\,N_{\rm e}=8.0$--10.0 and temperature range
$\log\,T=$5.65--5.95, the former calculated at 0.1~dex intervals, and
the latter at 0.05~dex intervals. The error on the theoretical value
is set to be the 3$\sigma$ varation of the ratio over the density and
temperature ranges.

\begin{deluxetable}{llll}
\tablecaption{Fe VIII insensitive ratios.\label{tbl.fe8.insens}}
\tablehead{& Ratio & Theory & Observation\tablenotemark{a}}
\tablewidth{0pt}
\startdata
Group 1 &187.24\tablenotemark{b}/186.60 &0.046 &$0.097\pm 0.004$\\
        &196.65\tablenotemark{c}/197.36 &0.151 &$0.498\pm 0.014$\\
        &193.97/194.66 &0.101 &$0.090\pm 0.004$ \btl \\
\noalign{\medskip}
%Group 2 &\lam185.21/\lam186.60  & $1.42\pm 0.23$ & $1.276\pm 0.016$  \\
%Group 2 &\lam195.97/\lam194.66  & $0.69\pm 0.07$ & $0.68\pm 0.01$ \\
Group 2  &\lam192.01\tablenotemark{d}/\lam197.36  & $0.215\pm 0.023$ & $0.398 \pm 0.019$ \\
        &\lam255.13/\lam253.98  & $0.561\pm 0.064$ & $0.395\pm 0.019$ \btl\\
        &\lam255.37/\lam253.98  & $0.556\pm 0.111$ & $0.678\pm 0.026$ \\
        &\lam255.71/\lam255.13  & $0.694\pm 0.099$ & $0.500 \pm 0.036$ \vtl\\

\noalign{\medskip}
Group 3 &\lam197.36/\lam194.66  & $0.385\pm 0.015$ & $0.422\pm 0.007$ \btl\\
        &\lam194.66/\lam185.21  & $0.191\pm 0.033$ & $0.305 \pm 0.004$ \btl\\
        &\lam197.36/\lam185.21  & $0.073\pm 0.013$ & $0.129\pm 0.002$ \btl
        \\
        &\lam206.75/\lam197.36  & $0.306\pm 0.070$ & $0.261\pm 0.021$ \\
\enddata
\tablenotetext{a}{A \vtl\ symbol indicates a $>1\sigma$ discrepancy
  between theory and observation, \btl\ indicates a $>2\sigma$
  discrepancy.}
\tablenotetext{b}{Blended with \ion{Fe}{vii} \lam187.235.}
\tablenotetext{c}{Blended with \ion{Fe}{vii} \lam192.006 and an
  unknown hotter line.}
\tablenotetext{d}{Blended with \ion{Fe}{xii} \lam196.640.}
\end{deluxetable}

\begin{deluxetable}{ll}
\tablecaption{New \ion{Fe}{viii} level energies.\label{tbl.fe8.energies}}
\tablehead{&Energy \\
Level & (cm$^{-1}$)}
\tablewidth{0pt}
\startdata
$3p^53d^2(^3F)$ $^4D^{\rm o}_{1/2}$  & 391\,115 $\pm$ 6 \\
$3p^53d^2(^3F)$ $^4D^{\rm o}_{3/2}$  & 391\,997 $\pm$ 6 \\
$3p^53d^2(^3F)$ $^4D^{\rm o}_{5/2}$  & 393\,463 $\pm$ 12 \\
$3p^53d^2(^3F)$ $^4D^{\rm o}_{7/2}$  & 395\,610 $\pm$ 12 \\
$3p^53d^2(^3P)$ $^2D^{\rm o}_{5/2}$  & 483\,671 $\pm$ 10 \\
\enddata
\end{deluxetable}

The Group 2 ratio \lam192.01/\lam197.36 involves lines emitted from the $3p^53d^2(^1S)$ $^2P$ term and Table~\ref{tbl.fe8.insens} shows it is discprepant with theory. This
is due to a blend of \lam192.01 with both a coronal line and a line
from \ion{Fe}{vii} (Sect.~\ref{sect.fe7}). Images formed in
the line suggest the blending line is probably \ion{Fe}{xi} and indeed
\citet{brown08} list the $3s^23p^4$ $^3P_1$ -- $3s^23p^3(^2D)3d$
$^3S_1$ transition from this ion. This identification is questionable,
though, given the inconsistent identifications from this upper level
given by \citet{brown08}: the decay to the ground $^3P_2$ level is
listed at both 187.45~\AA\ and 188.30~\AA, and the decay to the ground
$^1D_2$ level is listed at both 201.74~\AA\ and 202.70~\AA. Only the
187.45 and 201.74~\AA\ identifications are consistent with the
192.02~\AA\ line. The
CHIANTI model for \ion{Fe}{xi} has only a theoretical energy for the
$^3S_1$ level, with the decay to $^3P_1$ level listed at 191.21~\AA.
Using the \ion{Fe}{viii} \lam192.01/\lam197.36 theoretical ratio we
estimate a contribution of $41.5\pm 4.5$~\ecss\ of \ion{Fe}{viii} to the
measured 192.01~\AA\ intensity.  The remaining Group 2 ratios are
discussed later in this section with regard the newly-identified $3p^63d$ $^2D_{J}$ -- $3p^53d^2(^3F)$
$^4D_{J^\prime}^{\rm o}$ transitions. Note that the strong lines from
the  $3p^53d^2(^3F)$ $^2F$ and $3p^64d$ $^2P$ terms show temperature
and density sensitivity and so are not included in
Table~\ref{tbl.fe8.insens}. \lam185.21/\lam186.60 was discussed
earlier, while the ratio formed from lines from the $3p^64d$ $^2P$
term, \lam195.97/\lam194.66, has an observed ratio of $0.68\pm 0.01$
which is very close to the high density limit of the theoretical ratio
and implies a density
$\log\,N_{\rm e}\ge 8.64$, which is consistent with the \ion{Mg}{vii}
density value from Paper~I.

The Group 3 ratios in Table~\ref{tbl.fe8.insens} are formed from pairs
of lines belonging to different multiplets. None of the ratios agrees
with observations, although \lam197.36/\lam194.66 is only 10~\%\
discrepant. The discrepancies for \lam194.66/\lam185.21 and \lam197.36/\lam185.21 are 60 and
77~\%\ and suggest that the \ion{Fe}{viii} atomic model under-predicts
the strength of the \lam185.21 line. The remaining Group 3 ratio is in
better agreement with theory and is
discussed below.

The theoretical \ion{Fe}{viii} model in CHIANTI predicts eight
emission lines between 203 and 208~\AA, two of which are sufficiently
strong to be easily observable in the present EIS spectrum. Firstly we
note that the theoretical 
energies of \citet{griffin00} for the levels that give rise to the
lines in the 185--200~\AA\ range are all over-estimates of the
experimental energies. If we assume the energies for the levels that
give rise to the 203--208~\AA\ lines are also over-estimated by a
similar amount, then the lines should lie up to 4~\AA\ longward of the
theoretical wavelengths. The strongest lines predicted by CHIANTI are
the $3p^63d$ $^2D_{5/2}$ -- $3p^53d^2(^1G)$ $^2G_{7/2}^{\rm
  o}$ transition at 203.08~\AA\ and the $3p^63d$ $^2D_{3/2}$ -- $3p^53d^2(^3P)$ $^2D_{5/2}^{\rm
  o}$ transition at 205.01~\AA, whose strengths should be $0.31\pm 0.07$ and
$0.26\pm 0.05$ of the \lam197.36 \ion{Fe}{viii} line. A line at
206.78~\AA\ is the best match, in terms of intensity, to the
$^2G_{7/2}$ transition and we tentatively identify this transition. The image
formed in the line is consistent with other \ion{Fe}{viii} images, and
the Group~3 insensitive ratio, \lam206.75/\lam197.36, shown in
Table~\ref{tbl.fe8.insens} is consistent with theory. The new
experimental energy for the $3p^53d^2(^1G)$ $^2G_{7/2}$ level is given
in Table~\ref{tbl.fe8.energies}, and is derived assuming that a $+33$~\kms\
wavelength shift applies to the \ion{Fe}{viii} lines as discussed earlier in
this section. The error bar is derived using the measured centroid
uncertainty, the EIS wavelength scale uncertainty of $\pm 0.002$~\AA\
\citep{brown07}, and the uncertainty of the \ion{Fe}{viii} velocity
which is taken as the standard deviation of the six velocity
measurements of the unblended \ion{Fe}{viii} lines
(Table~\ref{tbl.fe8.wavelengths}). 
Two
lines at 208.68 and 208.84~\AA\ could be the $3p^63d$ $^2D_{3/2}$ -- $3p^53d^2(^3P)$ $^2D_{5/2}^{\rm
  o}$ transition, 
however the former has a significant  \ion{Cr}{viii} contribution
(Paper~I), while
the latter is too strong to be completely due to \ion{Fe}{viii}. We
thus do not make an identification for this transition in the
spectrum.

Four emission lines between 253.9 and 255.8~\AA\ that
are normally weak become very prominent in the current spectrum. Their
wavelengths and separations are close to those predicted by CHIANTI
for four lines of the \ion{Fe}{viii} $3p^63d$ $^2D_{J}$ -- $3p^53d^2(^3F)$
$^4D_{J^\prime}^{\rm o}$ multiplet, for which only theoretical wavelengths are
available, and images formed in the lines are very similar in
morphology to the \ion{Fe}{vii} and \ion{Fe}{viii} lines. We thus
identify the observed lines with the \ion{Fe}{viii} transitions. The
new experimental energy values for the $^4D_J$ levels are given in
Table~\ref{tbl.fe8.energies} which have been derived in the same
manner as the $3p^53d^2(^3P)$ $^2D_{5/2}$ discussed earlier. For the levels that decay
to the excited $^2D_{5/2}$ level in the ground configuration, the
$^2D_{5/2}$ energy of 1836~cm$^{-1}$  of \citet{ramonas80} has been
used. An uncertainty of $\pm$10~cm$^{-1}$ was assumed for this value.

Three ratios can be formed between the four lines that are relatively
insensitive to density and temperature, and  a comparison between
observation and theory is shown in Table~\ref{tbl.fe8.insens} (Group~2
lines).  Only one of the three ratios agrees with theory
within the error bars, but the disagreements are not large.

\begin{figure}[h]
\epsscale{0.5}
\plotone{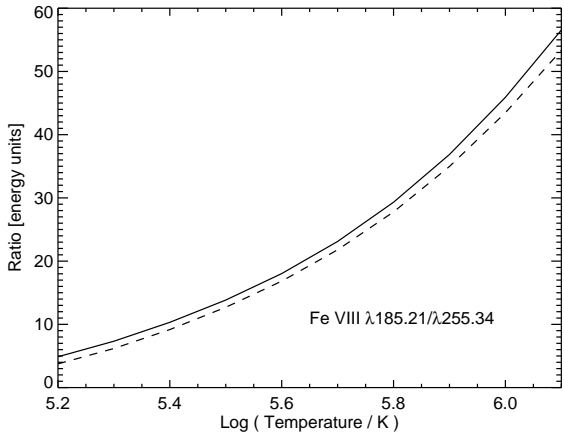}
\caption{Theoretical variation of the \lam185.21/\lam255.34 ratio as a
function of temperature. The ratio has been calculated for
densities of $10^9$~cm$^{-3}$ (solid line) and $10^8$~cm$^{-3}$
(dashed line). There is little density sensitivity above $10^9$~cm$^{-3}$. }
\label{fig.fe8.temp}
\end{figure}

All four lines are predicted to show significant temperature
sensitivity relative to the shorter wavelength lines between 185 and
200~\AA. In particular we highlight the \lam185.21/\lam255.34 ratio
which shows very little density sensitivity, and the temperature
variation is shown in Fig.~\ref{fig.fe8.temp}. However, the observed
ratio of $7.75\pm 0.23$ gives a very low temperature of
$\log\,T\approx 5.3$, well below the expected temperature of maximum
ionization of \ion{Fe}{viii} (Sect.~\ref{sect.ion-fraction}). This implies that the
\lam255.37 line is observed to be around a factor 3--4 stronger than
predicted by theory. It was noted earlier that insensitive ratios
amongst the SW lines suggest that the CHIANTI model is
under-estimating the strength of the \lam185.21 by 60--80~\%. If this
is the case, then the ratio curves plotted in Fig.~\ref{fig.fe8.temp}
  would be raised by this amount, making the derived \lam185.21/\lam255.37 temperature
even lower.

This raises the question of whether the
transitions between 254 and 256~\AA\ actually are \ion{Fe}{viii}
transitions. Firstly, we note that images formed in each line are
consistent with other \ion{Fe}{viii} lines. The line separations are
consistent with the theoretical energies of \citet{griffin00}, and the
observed wavelengths are within 1--2~\%\ of the theoretical
wavelengths. If the lines are not due to \ion{Fe}{viii}, then the
transitions must be nearby in the spectrum, but a similar set of four
lines with the correct intensities can not be found.
The only other ion species that we believe can be
responsible for the lines is \ion{Fe}{vii}, but the observed lines are
factors 5--10 stronger than the lines predicted from the 
\ion{Fe}{vii} atomic model in this wavelength range.

Our conclusion is thus that the observed lines are the \ion{Fe}{viii}
$3p^63d$ $^2D_{J}$ -- $3p^53d^2(^3F)$ 
$^4D_{J^\prime}^{\rm o}$ transitions but that the observed intensities
do not match the \ion{Fe}{viii} atomic model. To fix the discrepancy
would require the theoretical model to yield a greater level of
excitation to the $^4D$ levels. This can be achieved by cascading from
higher levels or increased resonance excitation, both of which could
be possible if the
$3p^43d^3$ configuration is included in the scattering calculation.

We summarise the results of the comparison between the \ion{Fe}{viii}
atomic model and the EIS spectrum as follows.

\begin{itemize}
\item The $3p^63d$ $^2D_{5/2}$ -- $3p^53d^2(^1G)$ $^2G_{7/2}^{\rm o}$
  has been identified at a rest wavelength of 206.753~\AA\, leading to
  a new experimental value for the $^2G_{7/2}$ energy.
\item Four transitions of the $3p^63d$ $^2D_{J}$ -- $3p^53d^2(^3F)$ 
$^4D_{J^\prime}^{\rm o}$ multiplet have been identified between 253.9
and 255.7~\AA, leading to new experimental energies for the four $^4D$
levels. The \ion{Fe}{viii} atomic model, however, underestimates the
lines' strengths by a factor between 3 and 6.
\item The two strong lines at 185.21 and 186.60~\AA\ emitted from the $3p^53d^2(^3F)$ $^2F$
  term are under-estimated by the atomic model by around 60--80~\%\
  compared to lines from the $3p^64d$ $^2P$ and
$3p^53d^2(^1S)$ $^2P$ terms. In addition, the ratio of the two lines is
discrepant with theory by around 10--20~\%.
\item The \lam185.21/\lam255.34 ratio is an excellent temperature
  diagnostic but yields temperatures significantly lower than expected
  due to the atomic data discrepancies highlighted above.
\item The group of four nearby lines at 193.97, 194.66, 195.97 and
  197.36~\AA\ show good agreement. \lam195.97/\lam194.66 could be a
  useful density diagnostic in conditions where the density is $\le
  10^9$~cm$^{-3}$. 
\end{itemize}

The discrepancies highlighted above suggest that a new atomic
calculation for \ion{Fe}{viii} is required. A new laboratory study of
\ion{Fe}{viii} would also be valuable for classifying the large number
of unidentified transitions between 200 and 260~\AA.

\section{Fe\,IX}\label{sect.fe9}

The CHIANTI 5 atomic model was revised following the new line
identifications of \citet{young09} and is made available in CHIANTI 6
\citep{dere09}. The new model is described in more detail below and is
compared with the \ion{Fe}{ix} 
lines in the 2007 February 21 spectrum. In addition we highlight a
number of observed lines that are likely to be due to \ion{Fe}{ix} but
for which definite transition identifications are not possible.

\begin{figure}[h]
\epsscale{1.0}
\plotone{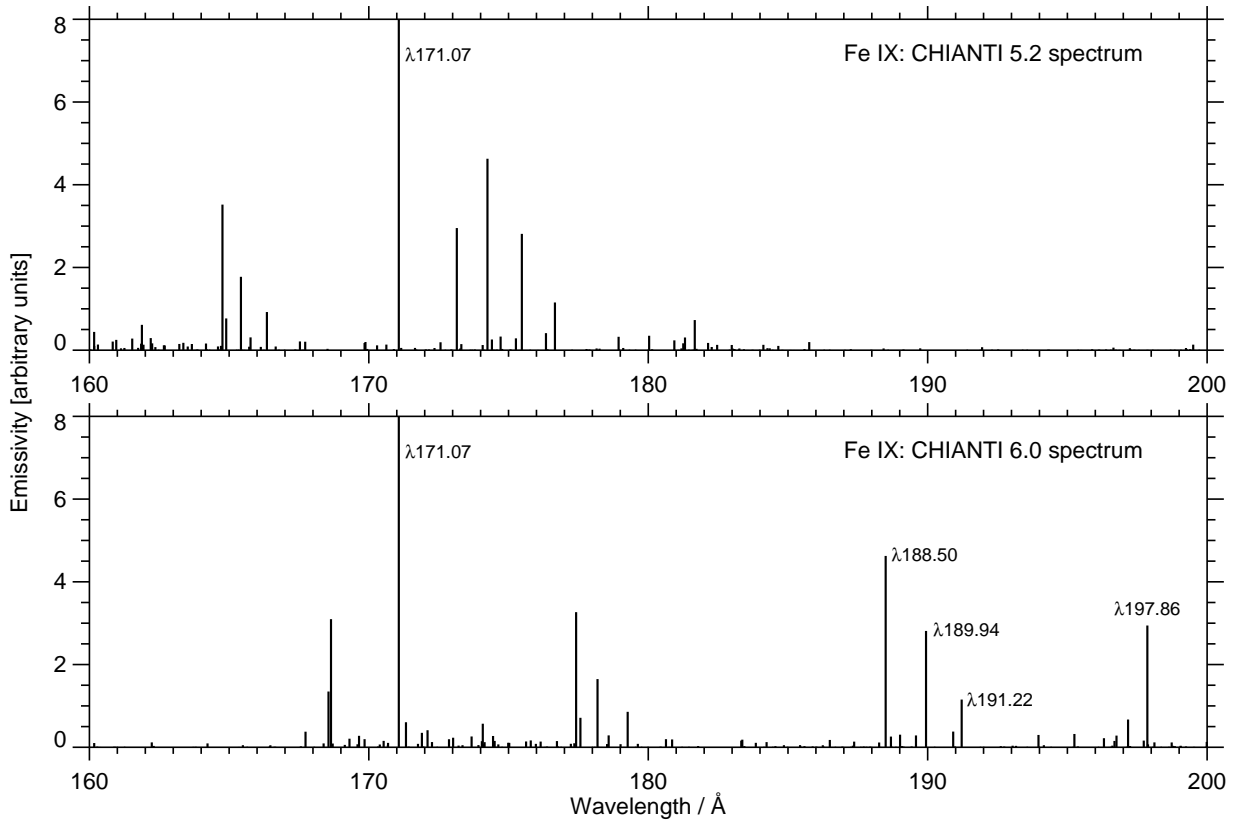}
\caption{Theoretical \ion{Fe}{ix} spectra calculated with CHIANTI 5.2
  (top panel) and CHIANTI 6 (bottom panel). Each emission line is
  represented by a vertical line, the length of which corresponds to
  the line's emissivity. The strong \lam171.07 line dominates,
  extending beyond the plots' bounds to Y-values of 98. The four lines
identified by \citet{young09} are indicated in the bottom panel.}
\label{fig.fe9-before-after}
\end{figure}

Four new \ion{Fe}{ix} line identifications were performed by
\citet{young09} and the energy levels for the ion have been updated for
CHIANTI 6 \citep{dere09}.
The $3p^4(^3P)3d^2$ $^3G_J$
levels were the first of the 109 levels of the $3p^43d^2$
configuration to be identified by any author, so they can be used to
provide energy corrections to the entire set of $3p^43d^2$
levels. These lead to improved wavelength estimates for other
transitions arising from the $3p^43d^2$ levels, a number of which are expected
in the EIS SW band. 
Since only one level multiplet has been identified, the average
difference between the experimental energies \citep[from][]{young09}
and theoretical energies \citep[from][]{storey02} has been calculated
and subtracted from each of the other fine structure levels of the
$3p^43d^2$ configuration. The energy subtracted is 43\,297~cm$^{-1}$,
and it shifts the predicted wavelengths by around 14~\AA\ compared to
the values in CHIANTI 5.2 \citep{landi06}. This method of adjusting the level energies
is very simplistic and the accuracy of the new wavelengths will not be
high (perhaps a few \AA), but they should be more accurate than the
previous estimates. The change in the \ion{Fe}{ix} wavelengths
following the energy shift is illustrated in
Fig.~\ref{fig.fe9-before-after}. Of particular interest are a strong triplet
arising from the $3p^4(^1D)3d^2$ $^3D_J$ levels between 177 and
180~\AA, and the group of weak lines between 188 and 199~\AA\ that
lie in a wavelength region where the EIS sensitivity is high. Before
considering these transitions, however, we discuss the
previously-identified transitions.

The strong \lam171.07 resonance line is found at the extreme short wavelength end of
the EIS SW band where the instrument effective area is very
low. Despite this, the line is well-resolved and, when converted to
calibrated intensity units, becomes almost a factor four stronger than
every other line in the spectrum.

The four lines identified by \citet{young09} are all found in the
present spectrum. The strongest line, \lam188.50, is partially blended with weak
lines in the short and long wavelength wings and is difficult to fit
accurately as there is no nearby continuum level. The fit parameters
shown in Paper~I resulted from a 
simultaneous 15 Gaussian fit to lines between 187.71 and 189.13~\AA. A
line in the short wavelength wing at 188.424~\AA\ is due to
\ion{Fe}{vii} and perhaps also a line 
of \ion{Mn}{ix} (see the discussion in Paper 1), while another
\ion{Fe}{vii} line is found in the long wavelength wing at
188.603~\AA. \lam189.94 is partially 
blended with \ion{Fe}{x} \lam190.04 but is easily resolved with a
double-Gaussian fit. \lam191.22 and \lam197.86 are both unblended.
The \lam197.86/\lam171.07 temperature diagnostic highlighted by
\citet{young09} yields a temperature of $\log\,T=5.86\pm 0.04$ which
is in agreement with the $T_{eff}$ value of 5.82.

The $3p^4(^3P)3d^2$ $^3G_{3,4}$ levels that give rise to the 
\lam189.94 and \lam191.26 lines also give rise to three weaker lines
that are potentially observable in the present spectrum. $^3G_4$
decays to $3p^53d$ $^3F_3$ with expected wavelength of
188.686~\AA, and the branching ratio is 0.092 relative to
\lam189.94. This line is observed in the spectrum with wavelength
$188.685\pm 0.007$~\AA, and the \lam188.686/\lam189.941 ratio is
$0.134\pm 0.034$, within $2\sigma$ of theory. There is a \ion{S}{xi}
line expected at 188.675~\AA\ 
which blends with the \ion{Fe}{ix} line, but using the calculated DEM
gives a predicted intensity of only 2.2~\ecss, therefore the
\ion{Fe}{ix} line dominates. An image formed in the 188.685~\AA\ line
confirms that it is a blend of a cool line around the temperature of
\ion{Fe}{ix} and a hotter line.

The $^3G_3$ level gives rise to \lam191.22 and has six additional
decays, two of which are 
potentially observable in the EIS spectrum. The strongest is the decay
to $3p^53d$ $^3F_3$ at
189.582~\AA, with a branching ratio of 0.248. It is a good wavelength
match to an observed line at 189.596~\AA, but the observed 
ratio is significantly lower than theory at $0.076\pm 0.016$. The
image formed in the line, however, is consistent with a formation
temperature close to \ion{Fe}{ix}.
The next strongest line from $^3G_3$ is the decay to $3p^53d$ $^1D_2$ at
199.986~\AA\ with a branching ratio of 0.112. This line is blended
with \ion{Fe}{xiii} \lam200.02 and contributes around 25~\%\ to the
observed feature (Paper~I). The image formed in this line is dominated
by the \ion{Fe}{xiii} component, however a weak cool component can be
identified, particularly if the image is formed in the short
wavelength wing of the line.

The $3p^53d$ $^3F_J$ -- $3p^4(^1D)3d^2$ $^3D_{J-1}$ ($J=4,3,2$)
triplet of lines is predicted at 177.419, 178.185 and
179.263~\AA, and are the strongest of the predicted lines between
\lam171.07 and \lam188.50 (see bottom panel of
Fig.~\ref{fig.fe9-before-after}). Although the lines are comparable in
strength to the longer wavelength triplet, the EIS effective area is
much lower around 177--180~\AA\ making the lines more difficult to
identify. 
 A good candidate for the stronger line is the observed
line at 176.968~\AA. It contains a \ion{Fe}{vii} component, but
Sect.~\ref{sect.fe7} demonstrated that this transition makes a
contribution of 30~\%. The remaining intensity of $313\pm 51$~\ecss\
is consistent with the theoretical \lam177.419 line: the
\lam177.419/\lam188.50 theoretical ratio is $\approx 0.7$, giving a
predicted intensity of $263$~\ecss\ using the measured \lam188.50
intensity.

With this identification the $^3F_3$--$^3D_2$ 
transition can be tentatively identified with the observed
177.603~\AA\ line based on wavelength separation and expected
intensity. The theoretical \lam178.185/\lam177.419 ratio is 0.51 for a
density of $10^9$~cm$^{-3}$ which, using the estimated intensity of
the measured 176.968~\AA\ line above, yields a predicted intensity of
160~\ecss\ in excellent agreement with the measured line intensity
(Paper 1). \citet{brown08} identified a line seen in quiet Sun and
active region spectra at this wavelength with \ion{S}{x} \lam177.545
and \ion{Ni}{xiv} \lam177.560, however, using the DEM distribution we can
demonstrate that these lines each contribute less than 1~\ecss\ to the
observed line at 177.603~\AA.
Based on wavelength separation and predicted intensity,
there are two candidates for the theoretical \lam179.263 line
($^3F_2$--$^3D_1$ transition): the 
measured lines at 178.708 and 178.994~\AA. The line is expected to
be a factor 0.27 weaker than \lam177.419 which is consistent with both
lines' measured intensities.  We note that the \ion{Fe}{ix} transition
$3p^53d$ $^3D_3$ -- $3p^4(^3P)3d^2$ $^3F_4$ 
has a predicted wavelength of 177.572~\AA\ and a predicted intensity
around 80~\%\ of \lam179.263 thus it could account for one of the two
observed lines.

In summary, we identify the observed 176.968 and 177.603~\AA\ lines
with the 4--3 and 3--2 components of the $3p^53d$ $^3F_J$ --
$3p^4(^1D)3d^2$ $^3D_{J-1}$ triplet, but do not make an identification
for the 2--1 component. New experimental energies for the
$3p^4(^1D)3d^2$ $^3D_{2,3}$ levels are given in
Table~\ref{tbl.fe9-energies}. The values have been derived by
adjusting the measured wavelengths by the average velocity shift of
the \lam188.50, \lam189.94, \lam191.22 and \lam197.86 lines
($+16$~\kms), and the uncertainties are a combination of the measured
wavelength uncertainties, the uncertainties of the four reference
wavelengths, and the uncertainties of the lower level energies
\citep[see][for more details]{young09}.
An updated theoretical energy for
the $^3D_1$ level is also given in Table~\ref{tbl.fe9-energies} and is
derived from the new $^3D_{2,3}$ energies by determining the average
difference in theoretical and experimental energies for these levels
and subtracting this from the theoretical $^3D_1$ energy.

\begin{deluxetable}{lll}
\tablecaption{New \ion{Fe}{ix} level energies.\label{tbl.fe9-energies}}
\tablehead{&&Energy \\
Level & Index\tablenotemark{a} & (cm$^{-1}$) }
\tablewidth{0pt}
\startdata
$3p^4(^1D)3d^2$ $^3D_3$ & 110 &$990913\pm 24$ \\
$3p^4(^1D)3d^2$ $^3D_2$ & 111 &$992393\pm 30$ \\
$3p^4(^1D)3d^2$ $^3D_1$ & 112 &993321\tablenotemark{b} \\
\enddata
\tablenotetext{a}{Index of the level in the CHIANTI 6 \ion{Fe}{ix} atomic model.}
\tablenotetext{b}{Energy derived from theoretical level splittings.}
\end{deluxetable}

No other \ion{Fe}{ix} lines can be definitively identified from the
EIS spectrum. However, Table~\ref{tbl.fe9-lines} gives a list of
measured lines that we believe are due to \ion{Fe}{ix} or have a
component due to \ion{Fe}{ix}. The lines have been identified through
comparisons of their image intensity distributions as described in
Paper~I, and also by comparing the Paper~I spectrum with that of
\citet{young09} which shows stronger \ion{Fe}{ix} emission relative to
\ion{Fe}{vii} and \ion{Fe}{viii} and so is valuable for discriminating
between these three ions for weak lines.

Table~\ref{tbl.fe9-predicted} gives a list of the six strongest
unidentified lines in the range 170--212~\AA\ which are thus the best
candidates for the suggested \ion{Fe}{ix} lines in
Table~\ref{tbl.fe9-lines}. The intensities are 
presented relative to \lam197.86, computed for a temperature of
$\log\,T=5.8$ and density $\log\,N_{\rm e}=9.0$. A short hand notation
is used so that $^3P_2$--$(^3P)^3D_3$, for example, corresponds to the
transition $3p^53d$ $^3P_2$ -- $3p^4(^3P)3d^2$ $^3D_3$. The strongest
of the six theoretical transitions is \lam197.174, and we note that a
further decay from this line's upper level is to the $3p^53d$ $^3F_4$
level with a branching ratio of 0.267. Identifying \lam197.174 with
the strongest of the observed lines at 194.816~\AA\ implies the decay
to $^3F_4$ occurs at 199.536~\AA\ which is close to the observed
199.613~\AA\ line. However, the difference in wavelength
(corresponding to 117~\kms) is too large given the accuracy of the
measured EIS wavelengths and the $3p^53d$ energy levels
\citep{edlen78}, and so we do not identify the observed 194.816~\AA\
line with the line of theoretical wavelength 197.174~\AA.

Identifying the \lam197.174 theoretical line with the observed line at
192.642~\AA\ would imply the decay to $^3F_4$ would occur at
197.255~\AA, however no observed line is found at this
wavelength. Finally if \lam197.174 was identified with the observed
line at 187.971~\AA, then the $^3F_4$ decay would be found at
192.361~\AA. This latter line would blend with \ion{Fe}{xii}
\lam192.39 and we note that a line is found in the short wavelength
wing of this line at 192.313~\AA, however this is 75~\kms\ away from the
expected position of the \ion{Fe}{ix} line and so the identification
can not be made.

Since no positive identification can be made for the strongest of the
six lines in Table~\ref{tbl.fe9-predicted}, then we do not attempt to
find identifications for the remaining lines. A new laboratory study
that can clearly separate \ion{Fe}{ix} lines  from the numerous
blending species found in the solar spectrum would be extremely
valuable for further classifying the many \ion{Fe}{ix} lines in the
 170--200~\AA\ wavelength range.

\begin{deluxetable}{lll}
\tablecaption{Possible \ion{Fe}{ix} emission lines.\label{tbl.fe9-lines}}
\tablehead{Wavelength & Intensity & \\
(\AA) & (\ecss ) & Class\tablenotemark{a}}
\tablewidth{0pt}
\startdata
186.004 & 18.5   & C--E \\
187.971 & 89.8   & E \\
192.642 & 76.7  & E--F \\
194.816 & 109.0 & C--E \\
195.753 & 20.2 & E,G--H \\
196.820 & 35.0 & E--G\\
199.613 & 34.4 & E--F \\
\enddata
\tablenotetext{a}{See Table~\tclass\ of Paper~I.}
\end{deluxetable}

\begin{deluxetable}{llll}
\tablecaption{Predicted \ion{Fe}{ix} intensities relative to \lam197.86.\label{tbl.fe9-predicted}}
\tablehead{ & Wavelength &  & Predicted intensity \\
Transition & (\AA) & Ratio\tablenotemark{a} & (\ecss )
 }
\tablewidth{0pt}
\startdata
$^3P_2$--$(^3P)^3D_3$ & 197.174 & 0.260 & 49 \\
$^3D_3$--$(^1S)^3F_4$ & 190.913 & 0.140 & 26 \\
$^3D_2$--$(^1S)^3F_3$ & 195.250 & 0.121 & 23 \\
$^1F_3$--$(^1D)^1G_4$ & 189.012 & 0.119 & 22 \\
$^3D_3$--$(^3P)^1G_4$ & 193.965 & 0.112 & 21 \\
$^1D_2$--$(^1D)^1F_3$ & 196.758 & 0.110 & 21 \\
\enddata
\tablenotetext{a}{Relative to \lam197.86.}
\end{deluxetable}

\section{Effects on response functions of EUV imaging instruments}

Solar EUV imaging instruments such as SOHO/EIT, TRACE
 and SECCHI/EUVI  employ multilayer
optical coatings to 
pick out narrow wavelength regions centered on specific EUV emission
lines. For example, each of these instruments has a channel centered
on the \lam195.12 line of \ion{Fe}{xii}. Although images formed in
this channel are dominated by \ion{Fe}{xii} and thus principally
reveal plasma 
at temperatures of 1.5~MK, many other emission lines within
$\approx$ $\pm$~10~\AA\ contribute to the channel and can modify the
response of the channel to temperature. For example, observations of
flares with TRACE \citep{gallagher02} have shown that the 195 channel becomes
dominated by \ion{Fe}{xxiv} \lam192.03 (formed at 20~MK) in the flare
region, while \citet{delzanna03} showed that lines of \ion{Fe}{viii},
\ion{Fe}{x} and \ion{Fe}{xi}
make a significant contribution to polar plume emission.

To determine the response of an EUV imaging channel to plasma
temperature it is necessary to compute synthetic spectra 
for a set of isothermal plasmas using an
atomic code. By convolving
the synthetic spectra with the instrument response function, one can
predict the instrument signal as a function of temperature. Since
different wavelength channels will have different responses to
temperature, then ratios formed from the channels (referred to as
filter ratios) can be used to determine the plasma temperature
\citep{moses97}. 
This method depends critically on the completeness of the atomic
models. For example, for the TRACE satellite the CHIANTI atomic
database was used to derive the instrument response function but,
prior to the version 3 release \citep{dere01}, CHIANTI contained no
atomic data for \ion{Fe}{viii} which led to the errors in the response
function for the 195 channel highlighted by \citet{delzanna03b}. 

For this work a new atomic atomic model for \ion{Fe}{vii} has been
prepared that predicts many new lines in the EUV that were not
previously included in CHIANTI. In addition, the \ion{Fe}{ix} model
has been revised significantly since the version~5.2 CHIANTI release
\citep{landi06} 
with new lines identified and many lines with theoretical wavelengths
shifted to new wavelengths. The present section investigates the
effects on response functions for the TRACE instrument.

The TRACE response functions currently in  Solarsoft were derived
using CHIANTI 5. To investigate the effects of the new \ion{Fe}{vii}
and \ion{Fe}{ix} models we use CHIANTI 5 with the \ion{Fe}{vii} and
\ion{Fe}{ix} data replaced with the new data. We use software developed by
\citet{brooks06} that derives the TRACE filter responses as
a function of temperature using the CHIANTI spectra. 

\begin{figure}[h]
\epsscale{1.0}
\plotone{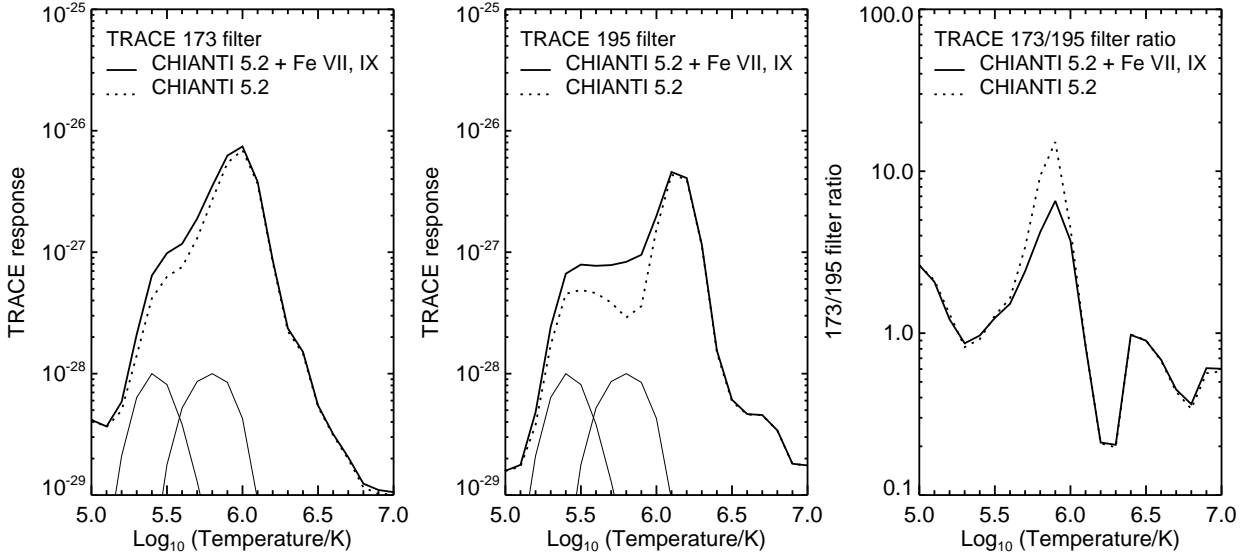}
\caption{The left and middle panels compare the response functions
  computed with CHIANTI 5.2 (dotted line) and CHIANTI 5.2 supplemented with
  the new \ion{Fe}{vii} and \ion{Fe}{ix} atomic data (solid line) for the 173 and 195
  channels, respectively. The functions are increased by the new data
  in the $\log\,T=5.5$--6.0 region, particularly for 195. At the bottom of
  both plots are shown the ionization fraction curves for Fe VII and
  Fe IX \citep[from][]{mazzotta98} to show their ranges of
  formation. The right panel plots the 173/195 filter ratio (which is
  used to derive plasma temperatures). The new curve is lower by up to
  a factor 3 over the region $\log\,T=5.5$--6.0.}
\label{fig.response}
\end{figure}

The left and middle panels of Fig.~\ref{fig.response} show the revised
response functions for the TRACE 173 and 195 channels. The 173 channel
has a greater sensitivity to \ion{Fe}{x} \lam174.53 than \ion{Fe}{ix}
\lam171.07 and so the curve peaks at the $T_{\rm max}$ of
\ion{Fe}{x}. The \ion{Fe}{ix} lines  between 170 and 180~\AA\ shown in
the lower panel of 
Fig.~\ref{fig.fe9-before-after} make a small increase to the response
function, particularly the $3p^53d$ $^3F_J$ -- $3p^4(^1D)3d^2$
$^3D_{J-1}$ ($J=4,3,2$) transitions discussed in
Sect.~\ref{sect.fe9}. Many \ion{Fe}{vii} lines between 170 and
180~\AA\ also increase the response function, with the $3p^63d^2$
$^3F_J$ -- $3p^53d^3(^4F)$ $^3F_{J}$ ($J=2,3,4$) transitions making the largest
contribution. 

The 195 channel response function shows a much greater change. In
particular a dip in the function that occurred at $\log\,T=5.8$
disappears due to the many \ion{Fe}{ix} lines now found between 188
and 200~\AA\ that were previously absent. The \ion{Fe}{vii} lines
between 195 and 197~\AA\ also make a significant increase to the
response function around $\log\,T=5.4$--5.7. 

Forming a ratio of the 173 and 195 channel response functions yields
the function shown in the right panel of
Fig.~\ref{fig.response}. Although significant structure is seen, the
most important feature is the fall between $\log\,T=5.8$ to 6.3 which
is the temperature range where \ion{Fe}{ix}, \ion{Fe}{x} and
\ion{Fe}{xii} are formed. Since in most coronal conditions these ions
dominate the channels' emission, then the slope means that the channel
ratio 173/195 can be used to diagnose temperatures in the range
$\log\,T=5.8$--6.3. This feature was first exploited for the SOHO/EIT
instrument \citep{moses97} which has very similar channels to TRACE,
and has been used in a number TRACE analyses, particularly studies of
coronal loops \citep{lenz99,aschwanden00}. Comparing the curves
derived with CHIANTI 5.2, with and without the new \ion{Fe}{vii} and
\ion{Fe}{ix} data, shows a significant change to the response function
ratio at $\log\,T=5.9$: the new ratio curve is a factor 2.3 lower than
previously. For $\log\,T\ge 6.0$, however, the change is $\le 20$~\%.

\section{Summary}

Eleven new line identifications have been performed for \ion{Fe}{vii},
leading to the assignment of eight new experimental energies for the
$3p^53d^3$ configuration, and the revision of two level assignments.
Four new line ratio diagnostics for temperature and density have been
highlighted, although discrepancies with values obtained through other
methods were found. Good agreement with
theory is found for around half of the \ion{Fe}{vii} lines, but there
are many discrepancies of up to a factor two. The problems are most
likely due to the atomic data for this complex ion, but a new
laboratory study of \ion{Fe}{vii} would be valuable, particularly for
confirming line identifications.

Five new line identifications have been performed for \ion{Fe}{viii},
leading to five new experimental energies in the $3p^53d^2$
configuration. Comparing observed line intensities with theory has
revealed a number of problems for the \ion{Fe}{viii} ion. The strong
transitions between 185 and 187~\AA\ arising from the
$3p^53d^2(^3F)$ $^2F$ term are not consistent with the  lines between
192 and 198~\AA\ emitted from the $3p^64d$ $^2P$ and $3p^53d^2(^1S)$
$^2P$ terms, being under-predicted by the atomic model by
60--80~\%. In addition, the newly identified multiplet from the
$3p^53d^2(^3F)$ $^4D$ term around 253--256~\AA\ is observed to be stronger than predicted
by theory by factors of between 3 to 6 compared to the lines in the
EIS SW wavelength band. A new atomic study would be valuable for
investigating these problems, while a new laboratory study of
\ion{Fe}{viii} between 170 and 260~\AA\ is required to identify the many
identified transitions in this range.

\citet{young09} identified the four strongest \ion{Fe}{ix} lines in
the EIS spectra and additional identifications have been performed
here. In addition, a number of emission lines have been classed as
\ion{Fe}{ix} lines based on image morphology but specific
atomic transitions could not be assigned. A laboratory study of
\ion{Fe}{ix} lines around 160--200~\AA\ is required to make further
progress. 

Solar EUV imaging instruments such as SOHO/EIT, TRACE, STEREO/EUVI and
the upcoming SDO/AIA require accurate atomic models for modeling the
dependence of the instruments' filters to plasma temperature. Using
the new atomic data and line identifications of \ion{Fe}{vii--ix} we
have modeled the TRACE 173 and 195 filter response functions,
demonstrating a significant change to the 195 filter response at
temperatures $\log\,T=5.4$--6.0.
This leads to a modification of the TRACE 173/195 filter ratio
that can affect coronal temperature determinations.

\acknowledgements

The work of EL is supported by the NNG06EA14I, NNH06CD24C and other NASA grants.
Hinode is a Japanese mission developed and launched by
ISAS/JAXA, with NAOJ as domestic partner and NASA and
STFC (UK) as international partners. It is operated by
these agencies in co-operation with ESA and NSC (Norway).

\end{document}